\documentclass{aa}

\usepackage{graphicx} 
\usepackage{txfonts}
\usepackage{hyperref}
\usepackage[switch]{lineno}
\usepackage{ulem}

\begin{document} 

\title{Three-dimensional spatial modelling of nuclear de-excitation lines}
\titlerunning{3D model of cosmic-ray de-excitation lines}

\author{
Tomohiko~Oka\inst{1} \fnmsep\thanks{Corresponding author: tomohiko.oka@uni-wuerzburg.de} 
\and
Thomas~Siegert\inst{1} 
\and
Tristan~Bouchet\inst{1}
\and
Laura~Eisenberger\inst{1}
\and
Mika~A.~Gelowicz\inst{1, 2}
\and
Saurabh~Mittal\inst{1}
\and
Rudi~Reinhardt\inst{1}
\and
Manuel~R.~H.~W.~Skalka\inst{1}
\and
Dimitris~Tsatsis\inst{1}
\and
Hiroki~Yoneda\inst{3}
\and
Manja~Zimmerer\inst{1}
}
\authorrunning{T. Oka et al.}
    
\institute{Julius-Maximilians-Universität W\"urzburg,
        Fakultät f\"ur Physik und Astronomie, Institut f\"ur Theoretische Physik und Astrophysik, Lehrstuhl f\"ur Astronomie,
        Emil-Fischer-Str.~31, D-97074 Würzburg, Germany
        \and        
        Rheinische Friedrich-Wilhelms-Universität Bonn, 
        Argelander-Institut f\"ur Astronomie (AIfA), 
        Auf dem H\"ugel 71, D-53121 Bonn, Germany
        \and
        Department of Physics, Kyoto University, 
        Kitashirakawa Oiwake-cho, Sakyo-ku, Kyoto 606-8502, Japan
        }

\date{March 2026}

 
\abstract
{Galactic cosmic rays (GCRs) are commonly thought to originate from shock acceleration in supernova remnants. Observations of ionization rates in the interstellar medium (ISM) indicate an excess of low-energy CRs (LECRs) beyond that expected from standard propagation models.
LECRs excite nuclei in the ISM, producing MeV $\gamma$-ray line emission through subsequent de-excitation. 
Observations of this emission can provide direct constraints on the LECR flux, its morphology, and its acceleration mechanisms.}
{We aim to construct all-sky maps of de-excitation $\gamma$-ray emission from the Milky Way based on three-dimensional (3D) CR distribution models. 
By comparing them with the Galactic diffuse $\gamma$-ray emission, we assess the detectability of the de-excitation line components against the underlying continuum.}
{We calculate the de-excitation emission by combining 3D CR distributions obtained with the CR propagation code GALPROP and $\gamma$-ray production cross sections calculated with the state-of-the-art nuclear reaction code TALYS.
Beyond the standard Galactic CR component, we construct models including an additional LECR component motivated by the high ionization rates measured in the ISM.}
{The predicted de-excitation emission is relatively weak when integrated over the broad regions commonly adopted in MeV $\gamma$-ray analyses ($|l|\leq30^{\circ}$ and $|b|\leq15^{\circ}$),
contributing only $\lesssim 1\%$ above the leptonic continuum even considering the additional LECR component.
The all-sky maps reveal that, in the standard scenario, the de-excitation-to-leptonic flux ratio remains low over the entire sky, reaching at most ${\sim}20\%$ toward the Galactic anticenter. In contrast, models including an additional LECR component exhibit a pronounced enhancement toward the GC, where the ratio exceeds $100\%$, indicating favourable prospects for detection through spatially resolved analyses.
We also find that the 6.1-to-4.4 MeV intensity ratio is particularly sensitive to the maximum energy of the injected LECR spectrum, varying by up to ${\sim} 30\%$.}
{Our results demonstrate that the next generation of MeV observatories can already establish the presence of a distinct LECR component and constrain its origin, acceleration mechanism, and transport in the Galaxy.}


\maketitle

\section{Introduction} \label{sec:introduction}
The origin of cosmic rays (CRs) has remained a mystery for more than a century \citep{Hess1912}.
Low-energy CRs below 1~GeV (hereafter, LECRs) are strongly affected by the heliospheric magnetic field, making it challenging to constrain their spectra at the source using only direct particle measurements within and around the heliosphere, including those obtained by Voyager \citep{Cummings2016ApJ}.
In the standard paradigm, CRs up to PeV energies are thought to be accelerated mainly by shock waves in supernova remnants (SNRs), but there is still no conclusive evidence for this scenario \citep[see, e.g.,][and references therein]{Blasi2013, Gabici2019IJMPD}.
From the viewpoint of astrochemistry, LECRs are considered to be a major source of ionization of hydrogen and helium in dense gas, and thus an important driver of chemical evolution in star-forming regions and protoplanetary disks \citep{Padovani2013A&A}.
However, measurements of the ionization rate toward the Galactic plane and the Galactic Center (GC) region indicate that the inferred LECR intensity is more than an order of magnitude higher than predicted by the standard CR propagation model, suggesting the presence of an additional LECR source \citep[e.g.,][]{Indriolo2009ApJ}.
As for the origin of the additional LECR component, several scenarios have been proposed, including astrosheaths \citep[the shocked stellar-wind regions between the termination shock and the astropause;][]{Scherer2008ApJ}, superbubbles \citep{Tatischeff2021MNRAS}, and  stochasticity effects in SNRs \citep{Phan2023PhRvD}.
However, the existence and properties of a distinct Galactic LECR component remain uncertain \citep[for a review, see][]{Gabici2022A&ARv}. 
Observational constraints on the LECR spectrum are therefore essential for establishing the presence of such a component and, if present, for investigating its origin and propagation in the Galaxy.

Two main indirect methods have been proposed to probe LECRs: one is to detect X-rays produced when LECRs ionize inner-shell electrons of atoms in the gas, followed by their de-excitation; the other is to detect $\gamma$-rays produced when LECRs excite atomic nuclei in the gas, followed by their nuclear de-excitation.
Observations using the former method have mainly been based on the 6.4\,keV line emission from neutral iron.
Several detections of 6.4\,keV line emission potentially associated with LECRs have been reported \citep{Tatischeff2012A&A, Nobukawa2015ApJ, Nobukawa2018ApJ}; however, since this line emission can be produced not only by LECR ions but also by X-rays and electrons, it is challenging to place stringent constraints on the LECR flux.
The latter method, based on MeV $\gamma$-ray observations, overcomes this limitation because the excitation processes are caused predominantly by LECRs.

To date, no LECR-induced MeV line emission has been detected from the Galactic plane, or even the GC.
Although a number of studies have predicted this $\gamma$-ray emission, the calculations have generally been limited to one- or two-dimensional spatial models and focused primarily on spectral predictions \citep[e.g.,][]{Indriolo2009ApJ, Dogiel2009A&A, Benhabiles2013ApJ, Liu2021A&A, Liu2024PRD}. 
In this study, we extend previous calculations to a three-dimensional framework and present the first skymap models of LECR-induced nuclear $\gamma$-ray line emission.
This skymap-based approach facilitates comparisons between nuclear line emission and leptonic continuum emission, the latter of which constitutes a major background for nuclear-line observations.
Physically motivated all-sky models are also crucial for comparisons with MeV $\gamma$-ray observations, since most current MeV instruments employ Compton-camera or coded-mask techniques, whose spectral analyses rely on assumed source distributions.
In addition, we investigate LECR source models motivated by ionization-rate measurements.
We construct LECR source models with parameters constrained by current CR observations and evaluate their expected $\gamma$-ray emission.

The structure of this paper is as follows. 
Section~\ref{sec:method} introduces the 3D modeling framework and the LECR source models investigated in this work.
Section~\ref{sec:results} presents the resulting sky maps and spectra of nuclear $\gamma$-ray line emission. 
Section~\ref{sec:discussion} discusses model uncertainties and the detectability of the predicted emission with current and future MeV observations.
Finally, Section~\ref{sec:summary} summarizes our main findings.

\section{Model framework and setup} \label{sec:method}

\subsection{Cross section data} \label{sec:method:cross_section}

We calculate cross sections for nuclear de-excitation line production using TALYS\footnote{\url{https://nds.iaea.org/talys/}} \citep[version 2.0;][]{Koning2023EPJA}, which predicts nuclear reactions based on state-of-the-art nuclear structure and reaction models.
Cross sections are calculated for proton- and $\alpha$-induced reactions over the energy range from 1\,MeV to 1\,GeV\footnote{
Although the recommended energy range for TALYS calculations is limited to 200\,MeV \citep{Koning2023EPJA}, the code has been extended to perform calculations up to 1\,GeV with reasonable agreement with experimental data demonstrated for total and non-elastic cross sections \citep{Koning2014}.
As shown in Sect.~\ref{sec:discussion}, uncertainties in the cross sections at high energies have only a minor impact on the resulting gamma-ray flux.
}.
TALYS calculates discrete-line, pre-equilibrium, and compound-nucleus emissions.
Following \citet{Murphy2009ApJS}, we classify 229 transitions as explicit lines.
The remaining transitions are used to construct a quasi-continuum component together with the pre-equilibrium and compound-nucleus emission components.
For this purpose, we exclude transitions above the particle-decay threshold and those violating spin--parity selection rules. 
The corresponding cross sections are summed into 100\,keV-wide energy bins to construct the quasi-continuum component.

\begin{figure}
\includegraphics[width=\linewidth,keepaspectratio]{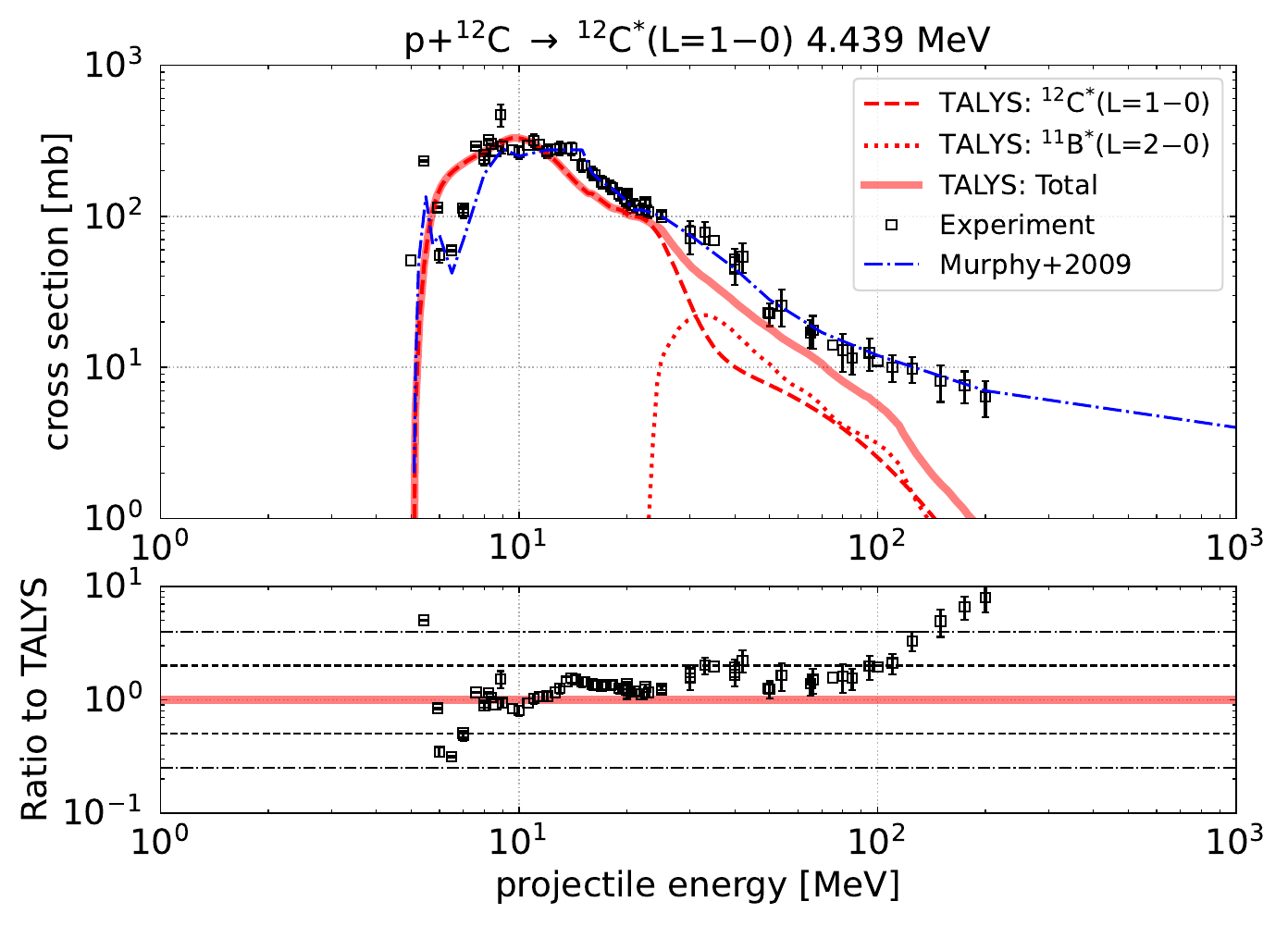}
\hfill
\includegraphics[width=\linewidth,keepaspectratio]{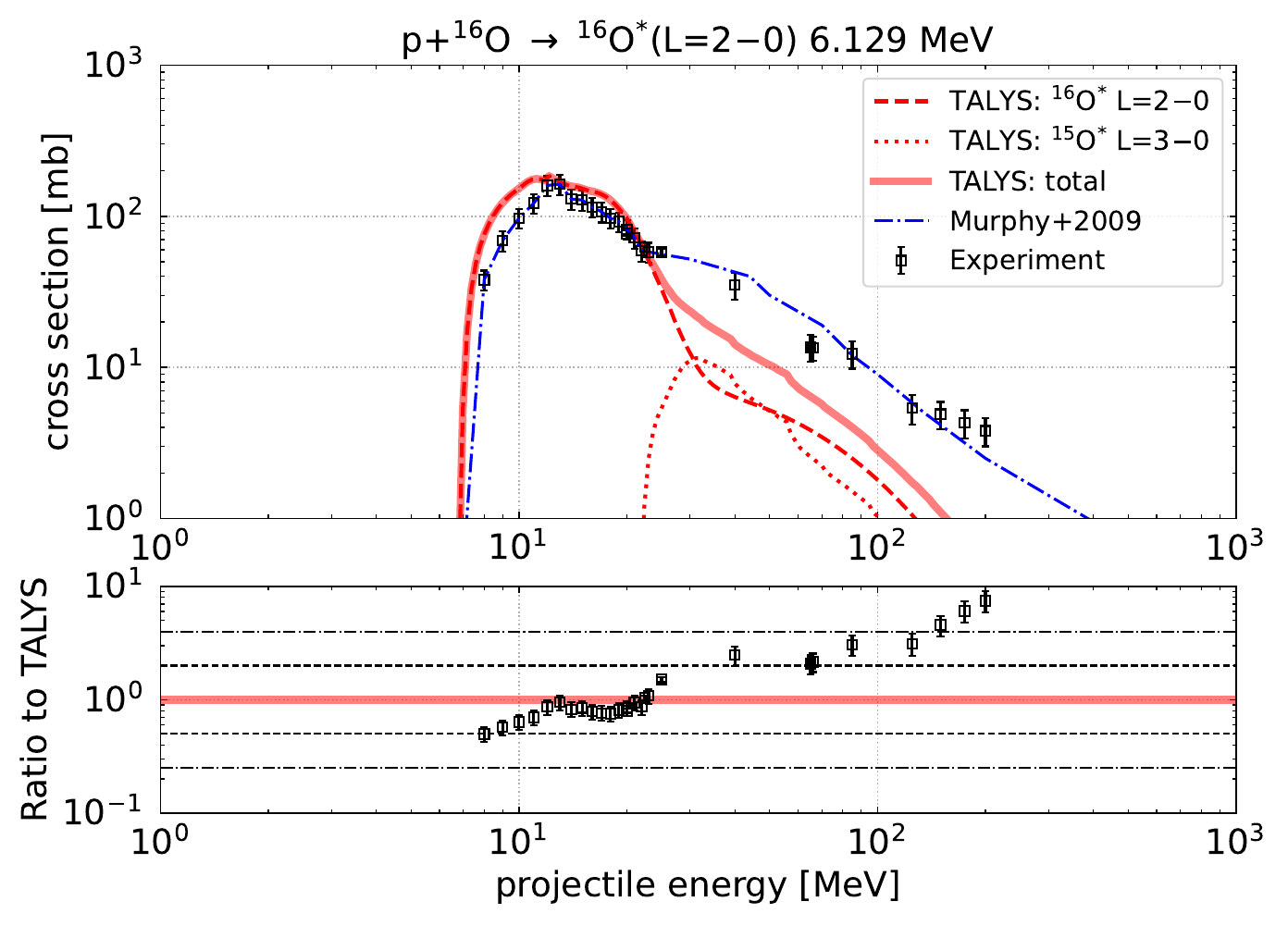}
\caption{Cross sections of $\gamma$-ray line emissions in proton-induced reactions.
The red curves are the results computed with TALYS \citep{Koning2023EPJA}, while the blue dot-dashed curves show the data provided by \citet{Murphy2009ApJS}.
The top and bottom panels correspond to the 4.439\,MeV line in p+$^{12}$C reactions and the 6.129\,MeV line in p+$^{16}$O reactions, respectively.
The data points show experimental data from the literature \citep{Dyer1981PhRvC, Lang1987PhRvC, Lesko1988PhRvC, Kiener2001PhRvC, Belhout2007PhRvC, Rahma2023NuPhA}.
}
\label{fig:csdata}
\end{figure}

Fig.~\ref{fig:csdata} shows the cross sections for the prominent 4.439\,MeV $\gamma$-ray line of $^{12}$C$^{*}$ produced in proton--$^{12}$C interactions, calculated with TALYS and compared with experimental data \citep{Dyer1981PhRvC, Lang1987PhRvC, Lesko1988PhRvC, Kiener2001PhRvC, Belhout2007PhRvC, Rahma2023NuPhA}. 
Since the data points shown include contributions from the nearby 4.445\,MeV line of $^{11}$B$^{*}$, we also include this transition when comparing with measurements above 20\,MeV, following \citet{Murphy2009ApJS}.
The resulting cross sections are in good agreement with the experimental data, generally within a factor of ${\sim}2$ below 100\,MeV.
A similar result is obtained for the 6.129\,MeV line. 
In this case, we include the contribution from the nearby 6.176\,MeV transition of $^{15}$O$^{*}$ when comparing the model predictions with the measurements.
We further confirm that the TALYS predictions are in reasonable agreement with the data for reactions with available experimental data.
This is consistent with the validation study of \citet{Benhabiles2013ApJ}, who reported agreement typically within a factor of ${\sim}2$.

It should be noted that all TALYS calculations presented here were performed using the default physics models and parameter values implemented in TALYS, without any user adjustment or optimization.
In several previous studies \citep[e.g.,][]{Benhabiles2011PhRvC,Rahma2023NuPhA}, the TALYS input parameters were optimized to better reproduce experimental cross-section data.
As shown in Fig.~\ref{fig:csdata}, calculations using the default TALYS settings can differ from the measurements by more than a factor of 2 at energies above 100\,MeV.
However, because experimental nuclear reaction data are available only for a limited number of reactions, such optimizations are applicable only to a subset of cases.
Therefore, in this study, we calculate the $\gamma$-ray flux using both the cross-section data obtained with the default TALYS settings and the measurement-based cross-section dataset compiled by \citet{Murphy2009ApJS}.
The differences between the resulting $\gamma$-ray fluxes are treated as a source of systematic uncertainty, and we later discuss the impact of the choice of cross-section data.

\subsection{Cosmic-ray model} \label{sec:method:cr_model}

\begin{figure}
\includegraphics[width=0.95\linewidth,keepaspectratio]{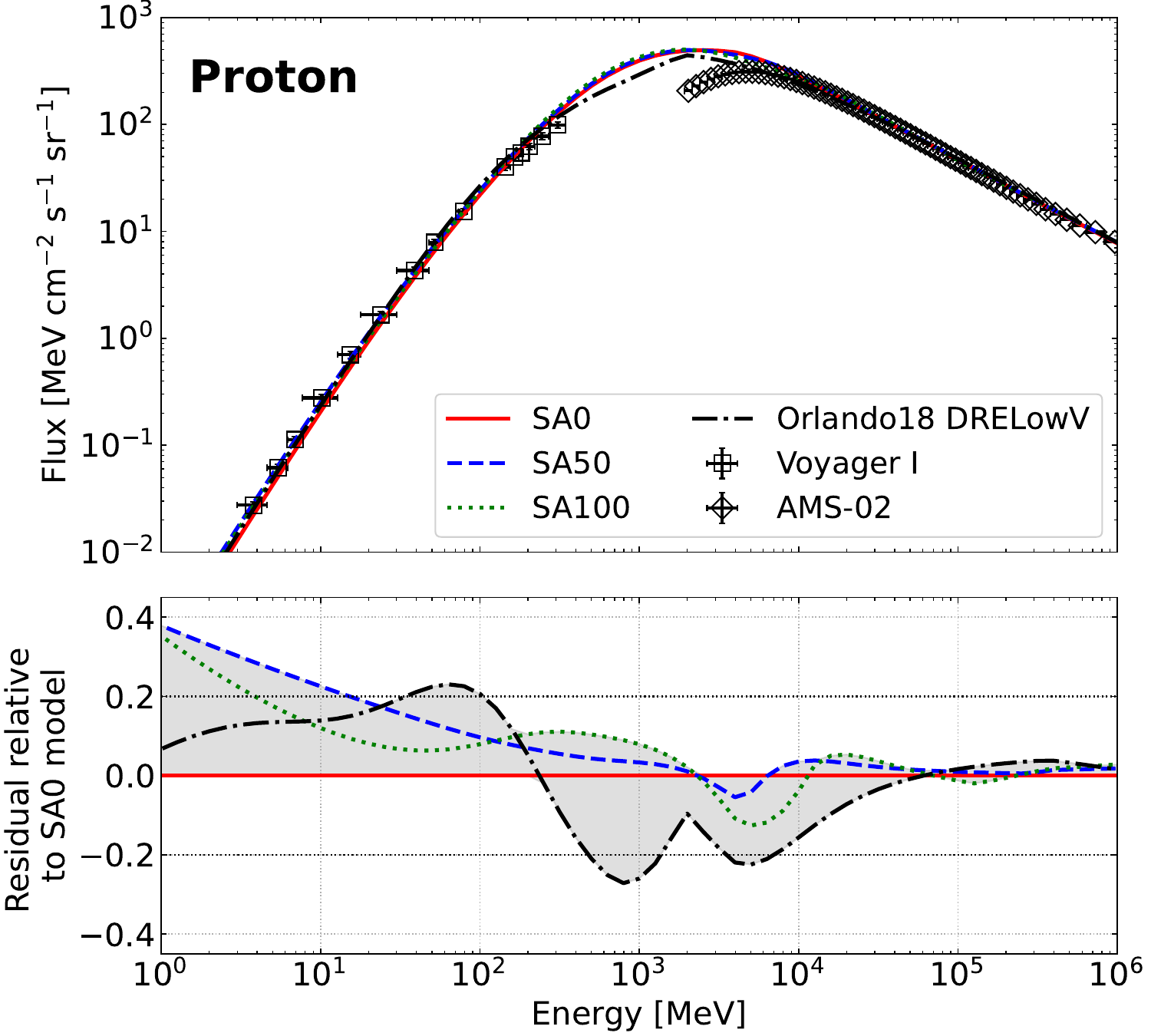}
\hfill
\includegraphics[width=0.95\linewidth,keepaspectratio]{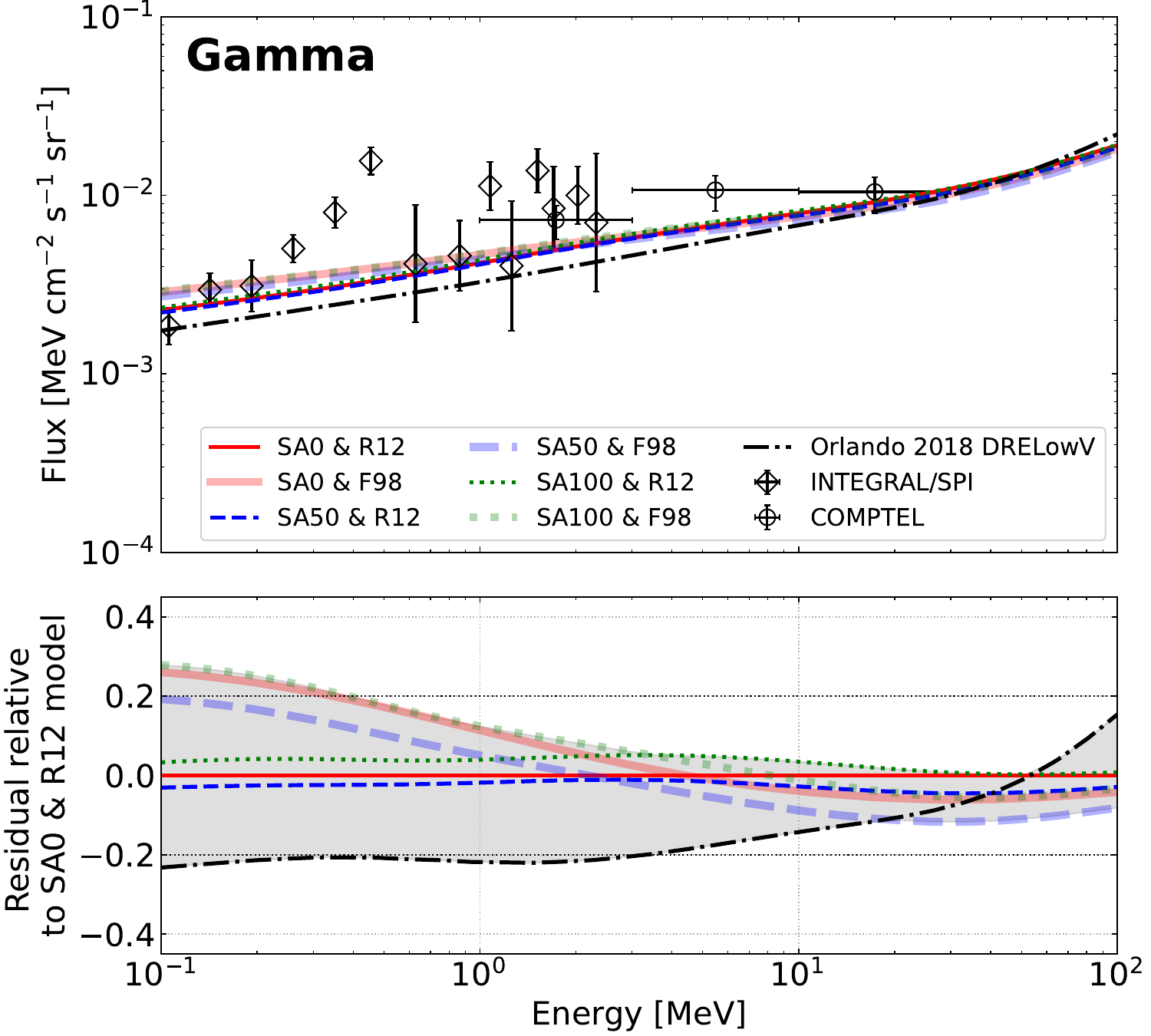}
\caption{
GALPROP model spectra obtained with the baseline parameter sets.
{\bf Upper panels:} Proton spectra near Earth and the corresponding residuals.
The red solid, blue dashed, green dotted, and black dot-dashed lines show the results for the SA0, SA50, SA100, and DRELowV models of \citet{Orlando2018MNRAS}, respectively.
Circle and square markers denote measurements from Voyager \citep{Cummings2016ApJ} and AMS-02 \citep{Aguilar2015PhRvL}, respectively.
Note that the deviation from the AMS-02 data below 10\,GeV is due to solar modulation, which is not considered here.
{\bf Lower panels:} Non-thermal $\gamma$-ray spectra integrated over the GC region, $|l|\le30^{\circ}$ and $|b|\le15^{\circ}$, and the corresponding residuals.
The colors and line styles are the same as in the upper panels.
Thick and thin lines denote the results obtained with the F98 and R12 interstellar radiation field models, respectively.
Diamond and circle markers denote measurements from INTEGRAL/SPI and CGRO/COMPTEL, respectively \citep{Bouchet2011ApJ}.
The residual panels show residuals with respect to the SA0+R12 model.
For clarity, only one set of results is shown in the upper spectral panel, as the proton spectra are independent of the adopted interstellar radiation field model.
}
\label{fig:galprop_baseline}
\end{figure}

To build the CR spectro-spatial distribution model, we employ GALPROP\footnote{\url{https://galprop.stanford.edu/}} \citep[version 57;][]{Porter2022ApJS}.
GALPROP numerically solves the transport equation of CRs, taking into account their injection, diffusion, decay, and energy losses.
Using GALPROP, we obtain 3D energy-dependent CR distributions for protons, $\alpha$ particles, and heavier nuclei ($^{12}$C, $^{13}$C, $^{14}$N, $^{16}$O, $^{20}$Ne, $^{22}$Ne, $^{24}$Mg, $^{26}$Mg, $^{27}$Al, $^{28}$Si, $^{32}$S, $^{40}$Ca, $^{56}$Fe; all nuclei included in the explicit-line table of \citet{Murphy2009ApJS}), together with the $\gamma$-ray spectra for each line of sight.
The latter include contributions from inverse Compton scattering, electron bremsstrahlung, and neutral-pion decay.
The GALPROP software provides several parameter sets that reproduce the observed CR and $\gamma$-ray spectra. 
For the 3D calculations, the parameters have been optimized for different combinations of CR source distributions (SA0, SA50, and SA100) and interstellar radiation field models (F98 and R12).
In addition, \citet{Orlando2018MNRAS} adjusted the electron injection spectrum, together with several propagation parameters, to reproduce the observed synchrotron emission.
Fig.~\ref{fig:galprop_baseline} shows proton and $\gamma$-ray spectra obtained with these parameter sets.
The resulting proton spectra show variations of typically 20--30$\%$ below 10\,GeV, while the $\gamma$-ray spectra likewise remain broadly consistent, with differences of ${\sim}20\%$.
Since the choice of the baseline model has only a minor impact on the results presented below, we adopt the SA0+R12 parameter set as our baseline model throughout this work.

\begin{figure*}
\includegraphics[width=\linewidth,keepaspectratio]{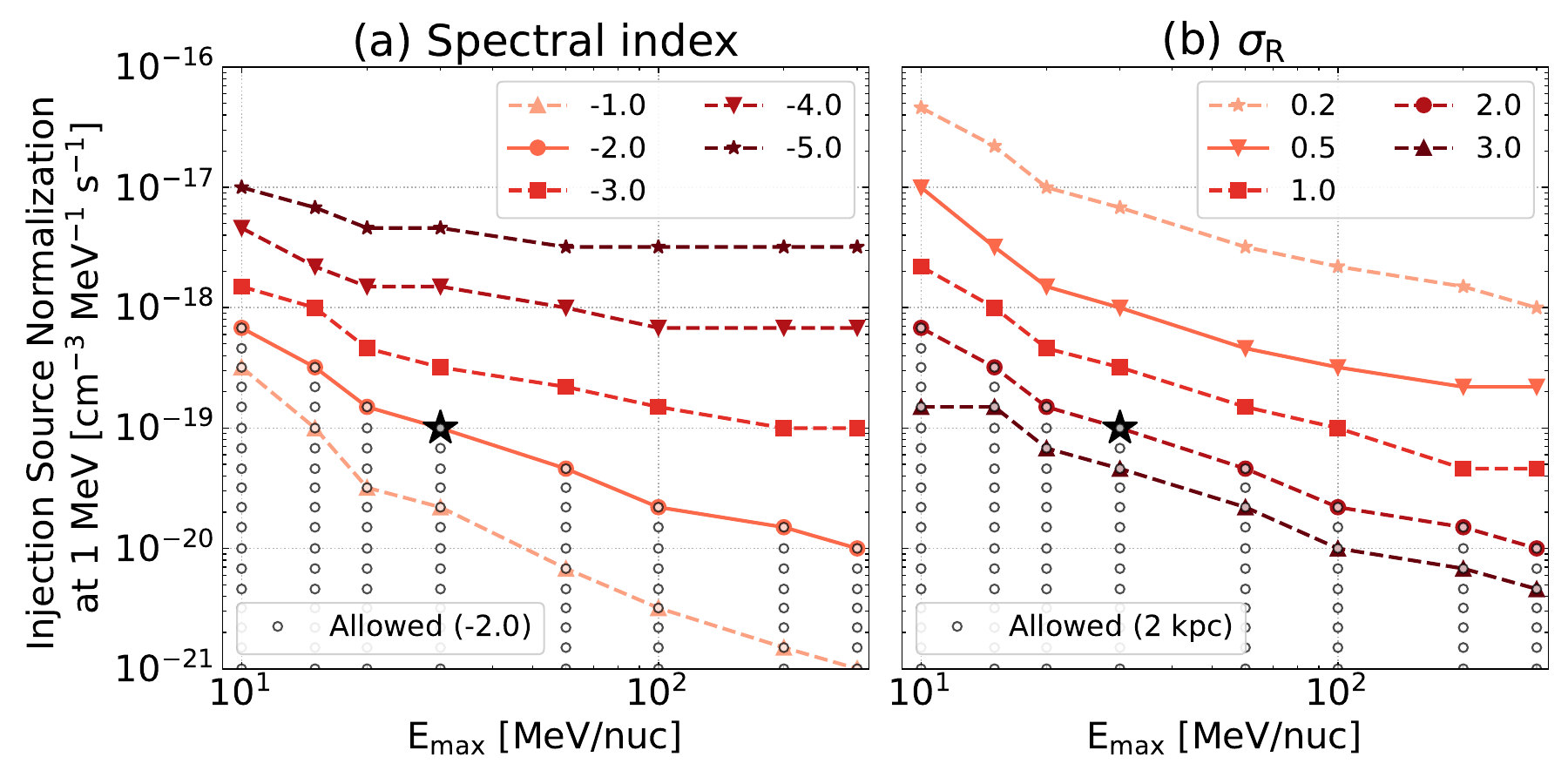}
\caption{Allowed LECR parameter region in the $N_{0}$--$E_\mathrm{max}$ plane.
The region to the lower left of each curve corresponds to the allowed parameter space.
Panel (a) shows the dependence on the injection spectral index for a fixed Gaussian width of $\sigma_{\rm R}=2$\,kpc, while panel (b) shows the dependence on $\sigma_{\rm R}$ for a fixed spectral index of $-2$.
In both panels, the black star indicates the fiducial parameter set.
}
\label{fig:lecr_parameter}
\end{figure*}

\begin{figure*}
\includegraphics[width=\linewidth,keepaspectratio]{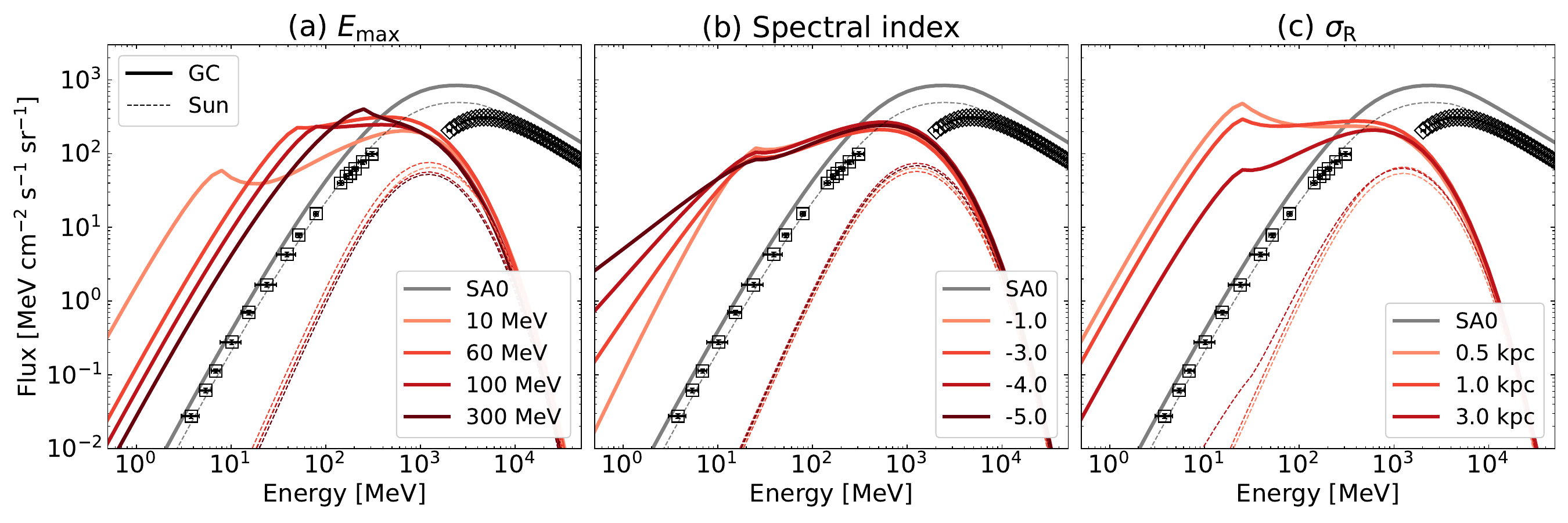}
\caption{
LECR spectra at the GC (R = 0\,kpc, Z = 0\,kpc; thick solid lines) and at the solar position (R = 8.5\,kpc, Z = 0\,kpc; thin dashed lines).
Panels (a)--(c) show the dependence on $E_\mathrm{max}$, the injection spectral index, and $\sigma_\mathrm{R}$, respectively.
For each parameter variation, the remaining parameters are fixed at the fiducial values, while the normalization ($N_{0}$) is set to the maximum value allowed by the observational constraints.
The gray curves show the baseline model (SA0).
}
\label{fig:lecr_spectra}
\end{figure*}

We also consider additional LECR components that are enhanced toward the GC.
As mentioned in Sect.~\ref{sec:introduction}, superbubbles and astrosheaths have been proposed as possible sites of LECR acceleration. Since both are associated with stellar populations, the corresponding LECR sources are expected to be concentrated along the Galactic plane and toward the GC. However, there is no established model for the spatial distribution of LECRs.
We therefore adopt a simplified model in which the LECR distribution is cylindrically symmetric, with a Gaussian radial profile and an exponential vertical profile:
\begin{equation}
    I = N_{0} \mathrm{exp} \left( - \frac{ R^{2} }{2 \sigma_\mathrm{R}^{2}} \right) \mathrm{exp} \left( - \frac{|Z|}{Z_{0}} \right),
\end{equation}
where $N_{0}$ is the injection source normalization, and $R$ and $Z$ denote the Galactocentric cylindrical radius and height above the Galactic plane, respectively.
This idea is similar to the bulge component inferred from 511\,keV observations \citep[e.g.,][]{Siegert2016A&A}, although in our model the radial extent ($\sigma_\mathrm{R}$) is allowed to vary.
The scale height ($Z_{0}$) is set to 0.2~kpc, following the baseline models; this value is compatible with the gas distribution \citep{Langer2014A&A}.
We assume that the injected LECR spectrum follows a truncated power-law in rigidity: $F(\mathcal{R}) = N_{0} \mathcal{R}^{s}$ for $\mathcal{R}_\mathrm{min}<\mathcal{R}<\mathcal{R}_\mathrm{max}$, where $s$ is the spectral index and $\mathcal{R}$ is the particle rigidity.
We choose the maximum rigidity for protons and $\alpha$ particles such that they correspond to the same maximum kinetic energy per nucleon, $E_{\rm max}$, and assume the same maximum rigidity as $\alpha$ particles for all heavier nuclei.
This treatment is broadly consistent with the assumption adopted by \citet{Benhabiles2013ApJ}, in which the maximum kinetic energy per nucleon is taken to be identical for all nuclei.
$E_\mathrm{min}$ is fixed at 0.2\,MeV\,nuc$^{-1}$ for all nuclei, corresponding to a particle velocity of about $6000$\,km\,s$^{-1}$ (equivalent to a temperature of $\sim2\times10^{9}$\,K for protons).

To select the LECR model parameters, we adopt a data-driven approach. 
We require that the inclusion of the additional LECR component does not significantly worsen the agreement between the baseline GALPROP model and the direct observed CR spectra.
Specifically, the additional LECR proton flux above 1\,MeV is required to remain below $20\%$ of the corresponding baseline flux, comparable to the variation among the baseline GALPROP models shown in Fig.~\ref{fig:galprop_baseline}.
We explore the parameter space spanned by $\sigma_\mathrm{R}$=0.2--3\,kpc, $-s$=$1$--$5$, and $E_{\rm max}$=10--300\,MeV\,nuc$^{-1}$.
Fig.~\ref{fig:lecr_parameter} shows the allowed parameter region.
As $E_{\rm max}$ increases, the LECR spectrum extends to higher energies, resulting in a larger flux around the spectral peak in $E^{2}dN/dE$ near the GeV range.
Therefore, the allowed source normalization becomes more tightly constrained.
Similarly, the normalization constraint becomes more stringent for larger $\sigma_\mathrm{R}$ and harder injection spectra.

As our fiducial model parameter set, we adopt $N_{0}$ = $10^{-19}$\,$\mathrm{cm^{-3}\,MeV^{-1}\,s^{-1}}$, s = $-2.0$, $E_{\rm max}$ = 30\,$\mathrm{MeV\,nuc^{-1}}$, $\sigma_\mathrm{R}$ = 2\,kpc.
We then explore the dependence on each model parameter by varying $E_\mathrm{max}$, $s$, and $\sigma_\mathrm{R}$ over the ranges considered in this study, while keeping the other parameters fixed at their fiducial values.
For each combination of $E_\mathrm{max}$, $s$, and $\sigma_\mathrm{R}$, the normalization ($N_{0}$) is set to the largest value in the allowed parameter region shown in Fig.~\ref{fig:lecr_parameter}.
Fig.~\ref{fig:lecr_spectra} shows the CR proton spectra for these representative cases.
We also refer to the model with $\sigma_\mathrm{R} = 0.2$\,kpc as the GC-concentrated case, which yields the highest $\gamma$-ray flux for the GC region ($|l| \leq 30^{\circ}$ and $|b| \leq 15^{\circ}$).
Note that, although the injected LECR spectrum is truncated at $E_{\rm max}\leq 300$\,MeV nuc$^{-1}$, the propagated spectrum extends to higher energies owing to diffusive reacceleration during CR transport \citep{Heinbach1995ApJ}.
Differences in the high-energy tail are relatively minor compared with those at lower energies.

\begin{figure}
\includegraphics[width=\linewidth,keepaspectratio]{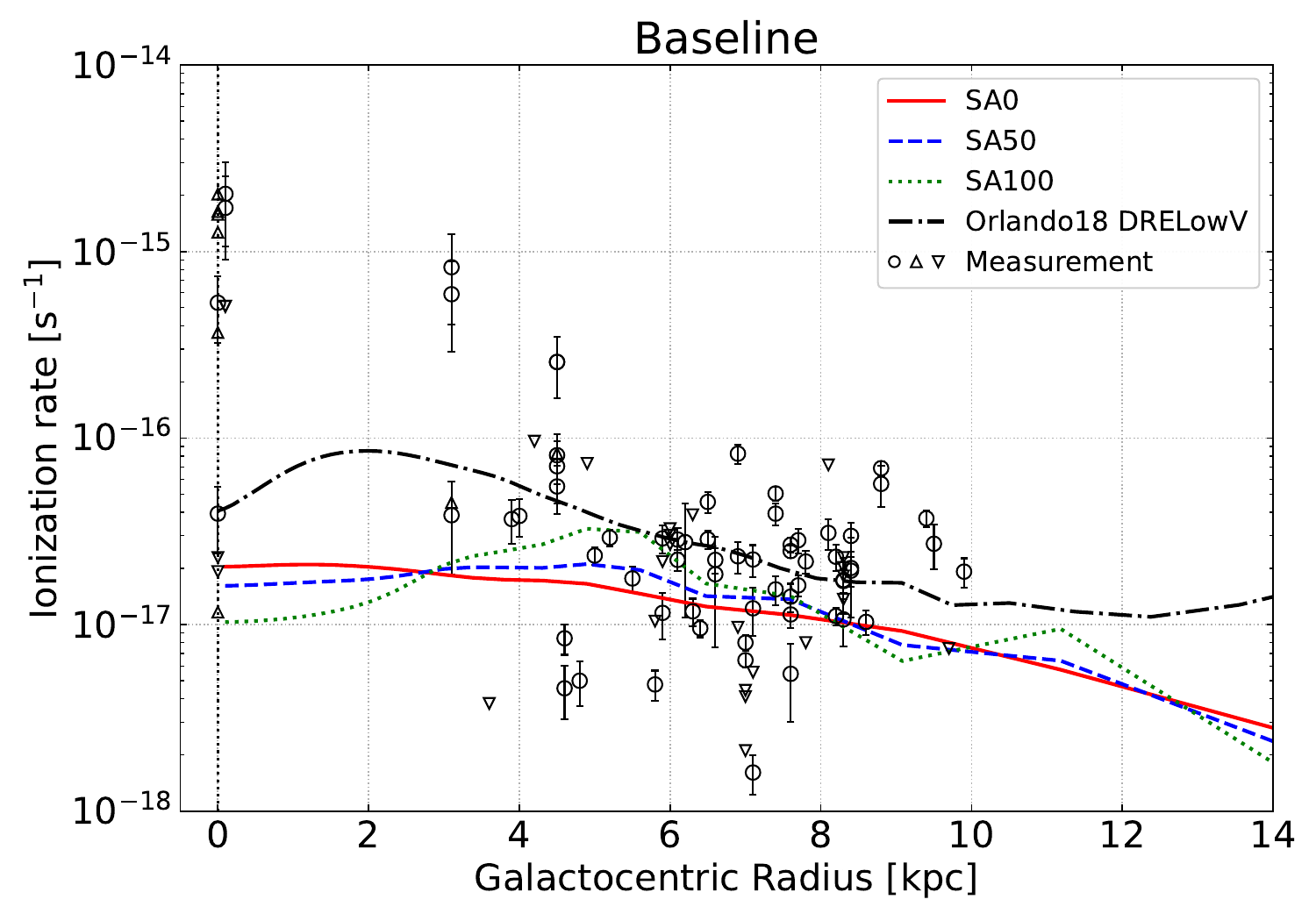}
\hfill
\includegraphics[width=\linewidth,keepaspectratio]{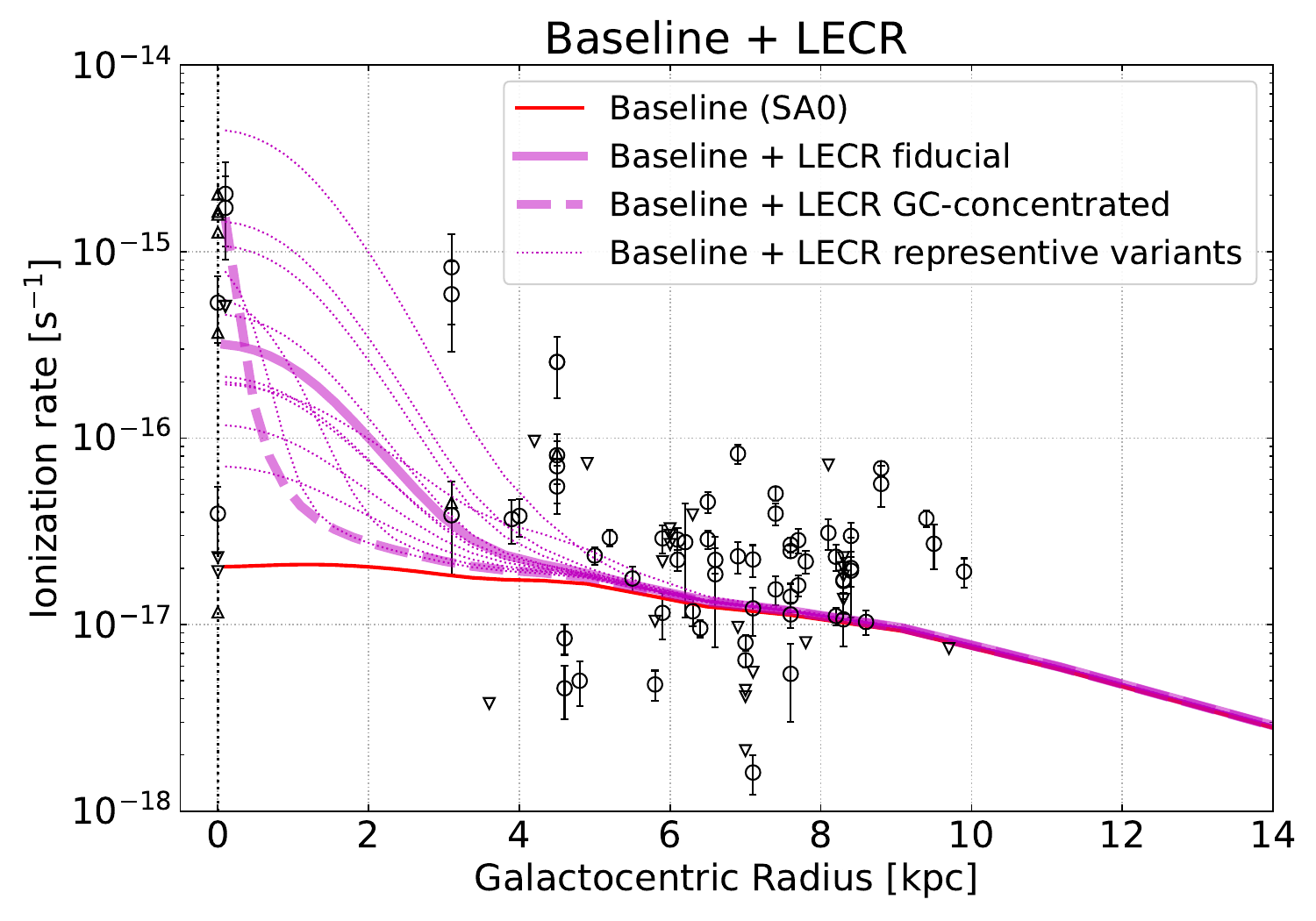}
\caption{Ionization rates as a function of Galactocentric radius. 
The circle markers denote measurements, while the upward- and downward-triangle markers indicate lower and upper limits, respectively, taken from \citet{Indriolo2015ApJ}, and scaled by a factor of 1/9, as suggested by recent reanalyses \citep{Obolentseva2024ApJ, Indriolo2026ApJ}.}
\label{fig:ionization_rate}
\end{figure}

Fig.~\ref{fig:ionization_rate} shows the ionization rates obtained with GALPROP together with the observational data from \citet{Indriolo2015ApJ}, scaled by a factor of $1/9$, as the recent reanalyses \citep{Obolentseva2024ApJ, Indriolo2026ApJ, Ho2026ApJ} have suggested that the ionization rates measured by \citet{Indriolo2015ApJ} may be overestimated by up to a factor of ${\sim}9$.
The results suggest that, despite the large variation among different regions
\footnote{
Recent JWST detections of CR-excited H$_{2}$ emission in the dark cloud Barnard~68 yield a high ionization rate of ${\sim}1.4 \times 10^{-16}\,\mathrm{s^{-1}}$ \citep{Bialy2026NatAs, Neufeld2026ApJ} at a Galactocentric radius of approximately 8\,kpc.
This finding provides further support for the possible presence of an additional LECR component and highlights the importance of a more detailed treatment of the spatial variations in the local CR distribution, which is beyond the scope of the present work, as discussed in Sect.~\ref{sec:discussion:remark}.
}, 
the baseline model can roughly reproduce the measurements at $\gtrsim 4$\,kpc but not toward the GC.
The additional LECR models, including the fiducial and GC-concentrated cases, may account for the enhanced ionization rates toward the GC.
Models with steeper spectral indices or lower $E_\mathrm{max}$ (and thus larger source normalization) predict higher ionization rates.
We further discuss the implications and limitations of the additional LECR model in Sect.~\ref{sec:discussion:remark}.

\subsection{Target material distribution} \label{sec:method:material}

We assume that the target material is traced by the Galactic H\,{\sc i} and CO distributions.
For consistency with the CR propagation calculations presented above, we adopt the gas density model implemented in GALPROP.
The number density distributions are parametrized as functions of the Galactocentric radius ($R$) and height ($Z$) following \citet{Moskalenko2002ApJ}.
For the line-of-sight integration, we further employ the GALPROP
H\,{\sc i} and CO Galactocentric gas ring maps\footnote{
We use \url{rbands_hi12_v2_qdeg_zmax1_Ts150_EBV_mag5_limit.fits.gz} and \url{rbands_co10mm_v3_2001_hdeg.fits.gz} as the templates for $N_{\mathrm{HI}}$ and
$W_{\mathrm{CO}}$, respectively.} to trace the spatial distribution of the interstellar gas
(see Eq.~\ref{eq:los_integration}).

For the elemental abundances in the ISM, we adopt the solar abundances of \citet{Lodders2009}. 
Although the metallicity is known to increase toward the GC and may reach 2--3 times the solar value in the innermost Galaxy \citep{Giveon2002A&A, Wenger2019ApJ}, we do not model the metallicity gradient. 
Instead, for simplicity, we assume a uniform metallicity of twice solar by doubling the abundances of all heavy elements (Z$>$2), following \citet{Benhabiles2013ApJ}.

\subsection{Line-of-sight integration} \label{sec:method:los_integration}

The $\gamma$-ray spectrum at each sky position is calculated by integrating the emissivity along the line of sight following the GALPROP formalism:
\begin{equation} \label{eq:los_integration}
    I_{\gamma} = \sum_{i} \frac{N_{\mathrm{HI}, i} + 2 X_{\mathrm{CO}, i} W_{\mathrm{CO}, i}}{\int_{\mathrm{ring}, i} (n_\mathrm{HI} + 2 n_\mathrm{CO}) ds} \int_{ring, i} q_{\gamma}  \, ds,
\end{equation}
where $N_{\rm HI}$ and $n_{\rm HI}$ are the H\,{\sc i} column and number densities, respectively, $W_{\rm CO}$ is the integrated CO intensity, $X_{\rm CO}$ is the CO-to-H$_{2}$ conversion factor, and $q_\gamma$ is the $\gamma$-ray emissivity.
The emissivity is given by
\begin{equation}
    q_{\gamma} = (n_\mathrm{HI} + 2 n_\mathrm{CO}) R_\mathrm{mass} \int \sigma F_\mathrm{CR} dE,
\end{equation}
where $R_\mathrm{mass}$ is the abundance of the target nucleus in the ISM relative to hydrogen, $\sigma$ is the $\gamma$-ray production cross section, and $F_\mathrm{CR}$ is the differential CR flux.
The integration is performed up to a distance of 30\,kpc with a step size of 0.1\,kpc along each line of sight.

We consider both explicit-line and quasi-continuum emission, where the explicit-line component includes both direct and inverse reactions.
Direct reactions arise from the excitation of interstellar heavy nuclei by CR protons and $\alpha$ particles, whereas inverse reactions are produced by interactions of heavy CR nuclei with interstellar H and He. 
The $\gamma$-ray lines from inverse reactions are thus broadened by the relativistic Doppler effect \citep[e.g.,][]{Huang2010PhRvA, Yoneda2023MNRAS}, and the line profiles are calculated by averaging the corresponding Doppler-broadened response functions over CR energy weighted by $\sigma F_{\rm CR}$.
In direct reactions, the de-excitation lines are Doppler-broadened by the recoil of the excited target nuclei.
For example, the intrinsic width of the 4.4\,MeV line is ${\sim}100$\,keV for isotropic interactions of incident protons \citep{Kiener2001PhRvC}.
This prediction is comparable to the line widths observed in solar flares, where the 4.4 and 6.1\,MeV lines have full widths at half maximum (FWHMs) of 100--150\,keV, while the 1.63\,MeV line has a FWHM of ${\sim}20$\,keV \citep{Smith2003ApJ, Harris2007A&A}.
For simplicity, we do not explicitly include recoil broadening in the present calculations.
Instead, we adopt 300 logarithmically spaced $\gamma$-ray energy bins spanning from 100\,keV to 100\,MeV, yielding energy-bin widths comparable to the expected recoil broadening. 

\section{Results} \label{sec:results}

\subsection{Spectra from conventional large-scale regions} \label{sec:results:large_scale}

\begin{figure*}
\centering
\includegraphics[width=\linewidth]{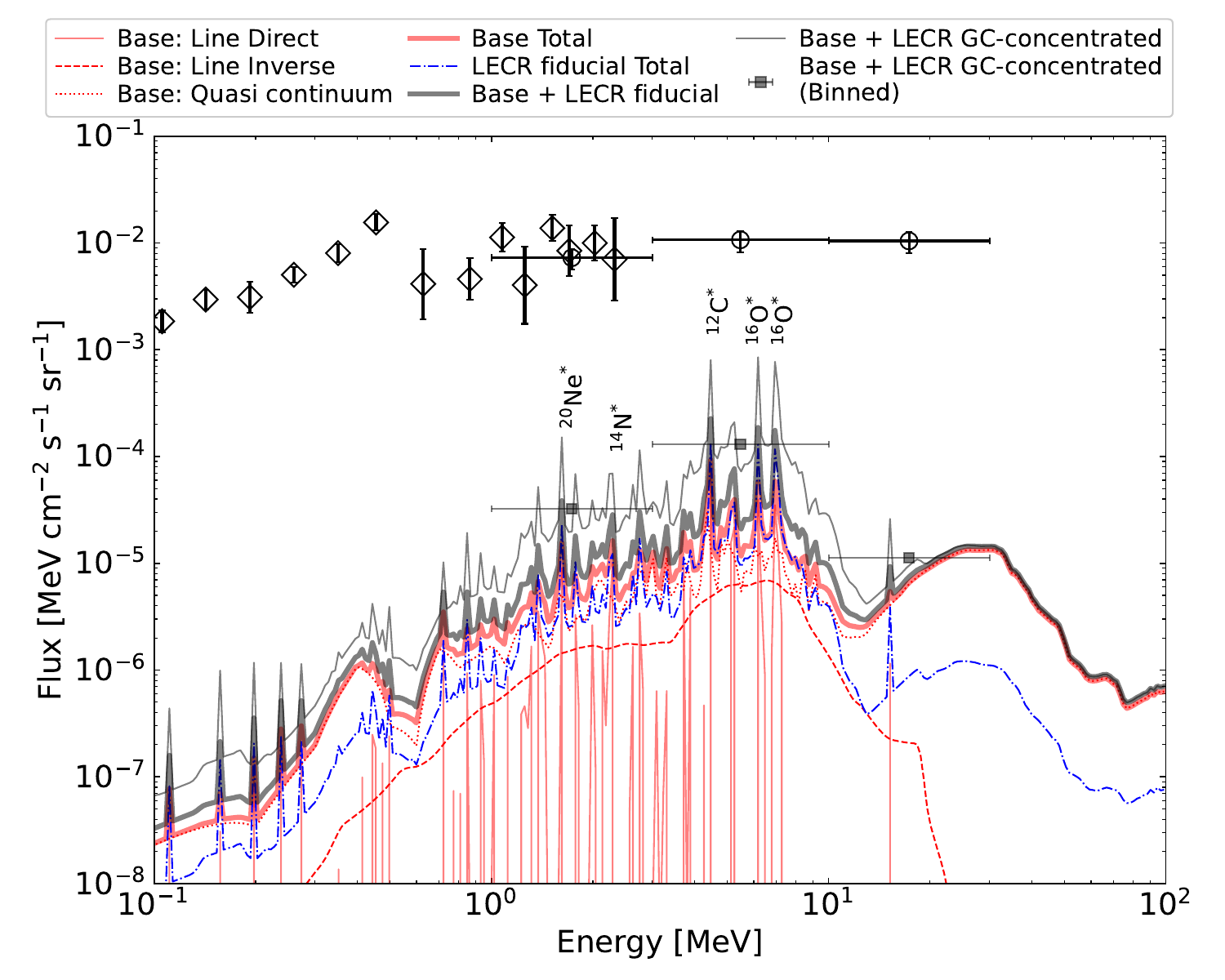}
\caption{
De-excitation $\gamma$-ray spectra of the inner Galactic ridge ($|l|\le30^{\circ}$ and $|b|\le15^{\circ}$). 
The red thin-solid, dotted, dashed, and thick-solid curves show the baseline model contributions from direct reactions, inverse reactions for explicit lines, quasi-continuum emission, and their total, respectively.
The blue dot-dashed curve shows the total spectrum of the fiducial LECR model, while the black thick-solid curve shows the combined spectrum of the baseline and fiducial LECR models.
The black thin-solid curve corresponds to the GC-concentrated LECR model, in which $\sigma_\mathrm{R}=0.2$\,kpc instead of the fiducial value of 2\,kpc. 
For the combined baseline and GC-concentrated LECR model, we also show the spectrum rebinned to the COMPTEL energy bins between 1 and 30\,MeV as thin black squares.
The other markers indicate the measured fluxes and are identical to those shown in the lower panel of Fig.~\ref{fig:galprop_baseline}.
}
\label{fig:spectrum_base}%
\end{figure*}

Fig.~\ref{fig:spectrum_base} shows the de-excitation $\gamma$-ray spectrum for the baseline model and the fiducial LECR model.
The predicted $\gamma$-ray spectrum exhibits prominent line features at 4.4\,MeV from $^{12}$C$^{*}$ and 6.1\,MeV from $^{16}$O$^{*}$, as well as several distinct lines, e.g., at 1.6\,MeV ($^{20}$Ne$^{*}$), 2.3\,MeV ($^{14}$N$^{*}$), and 6.9\,MeV ($^{16}$O$^{*}$).
Integrating over the same region adopted by \citet{Benhabiles2013ApJ} ($|l| \le 60^{\circ}, |b| \le 10^{\circ}$), the line intensities of the 4.4, 6.1, and 1.6\,MeV lines in the baseline model are $5.8 \times 10^{-7}$, $2.6 \times 10^{-7}$, and $2.7 \times 10^{-7}$\,$\mathrm{ph\,cm^{-2}s^{-1}sr^{-1}}$, respectively.
These values correspond to approximately 27--40$\%$ of those obtained in the standard case of \citet{Benhabiles2013ApJ}, mainly because inverse reactions, which account for ${\sim}70\%$ of the line flux in that study, contribute much less in the present calculation owing to their broader top-hat-like profiles, which are consistent with the recent study of \citet{Liu2021A&A}, which employed the traditional code of \citet{Ramaty1979ApJS}.
Note that the treatments of the cross-section data also differ among these studies.
\citet{Liu2021A&A} directly adopted the semi-empirical cross sections compiled by \citet{Murphy2009ApJS}, whereas \citet{Benhabiles2013ApJ} incorporated more recent measurements \citep{Benhabiles2011PhRvC}.
These differences could potentially affect the results by up to a factor of 2 (see Sect.~\ref{sec:discussion:cross_section}).

Compared with the baseline model, the line intensities increase by a factor of 1.4--2.2 in the fiducial LECR model and increase further for models with smaller values of $\sigma_{\rm R}$.
For $\sigma_{\rm R}=0.2$\,kpc, the predicted line intensities are enhanced by approximately an order of magnitude relative to the baseline model.
The dependence of the line intensities on the model parameters is discussed in more detail in Sect.~\ref{sec:results:line_ratio}. 
We note that, even after including the additional LECR component, the predicted de-excitation flux amounts to $\lesssim 1\%$ of the leptonic continuum in the entire Galactic plane and therefore is unlikely to explain the excess $\gamma$-ray emission toward the GC region, commonly referred to as the ``COMPTEL excess'' \citep[e.g.,][]{Strong2000ApJ, Porter2008ApJ, Bouchet2011ApJ}.

We also extracted $\gamma$-ray spectra for several additional integration regions where observational results are available and confirmed that our model predictions are consistent with the non-detection of de-excitation lines in all cases.
The details are presented in Appendix~\ref{sec:app:spectrum_vs_obs}.

\subsection{All-sky maps and region-specific spectra} \label{sec:results:skymap}

\begin{figure*}
\includegraphics[width=0.48\linewidth,height=0.25\textheight,keepaspectratio]{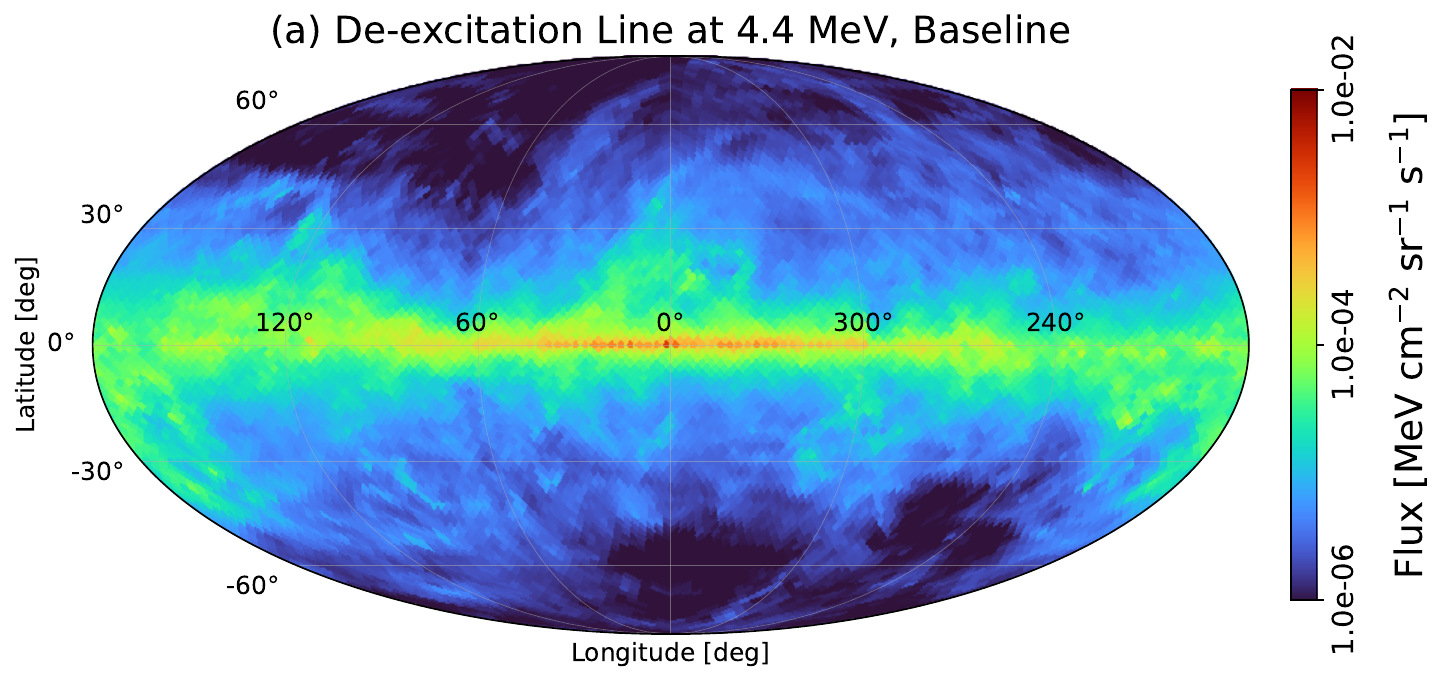}
\hfill
\includegraphics[width=0.48\linewidth,height=0.25\textheight,keepaspectratio]{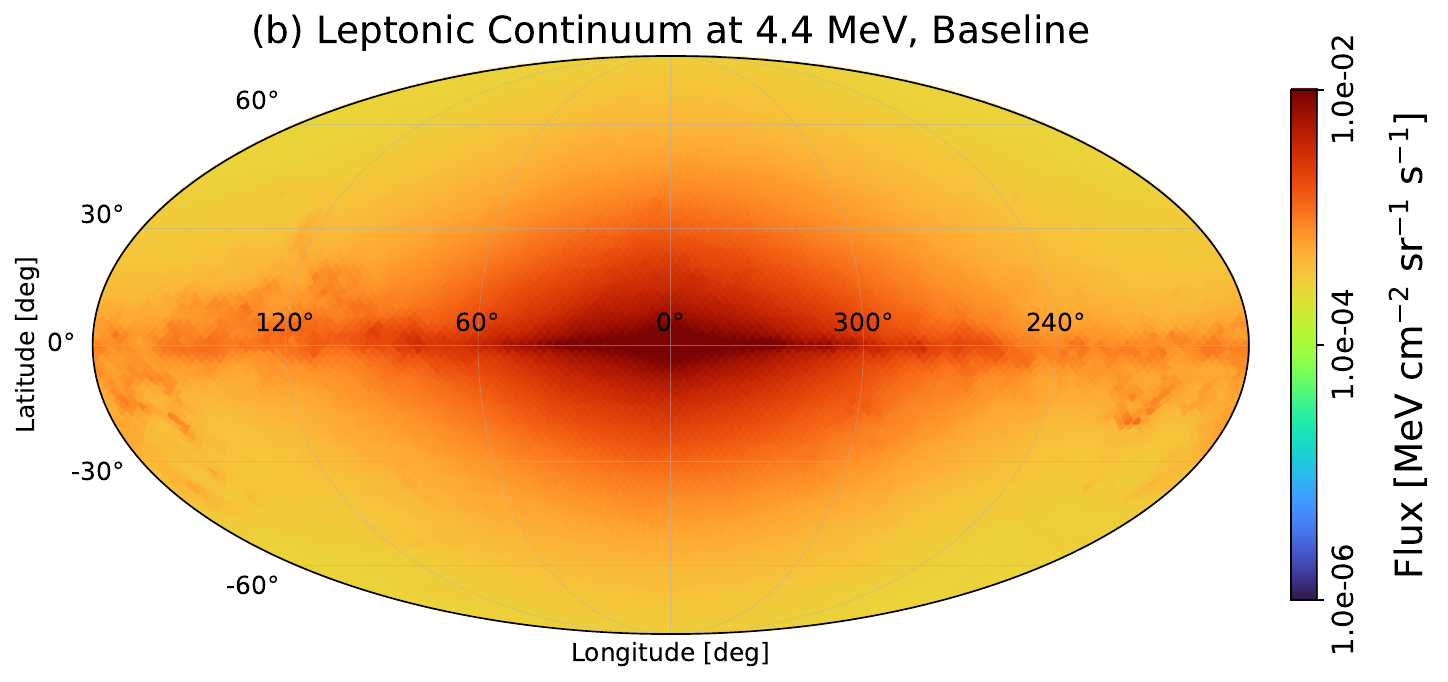}
\includegraphics[width=0.48\linewidth,height=0.25\textheight,keepaspectratio]{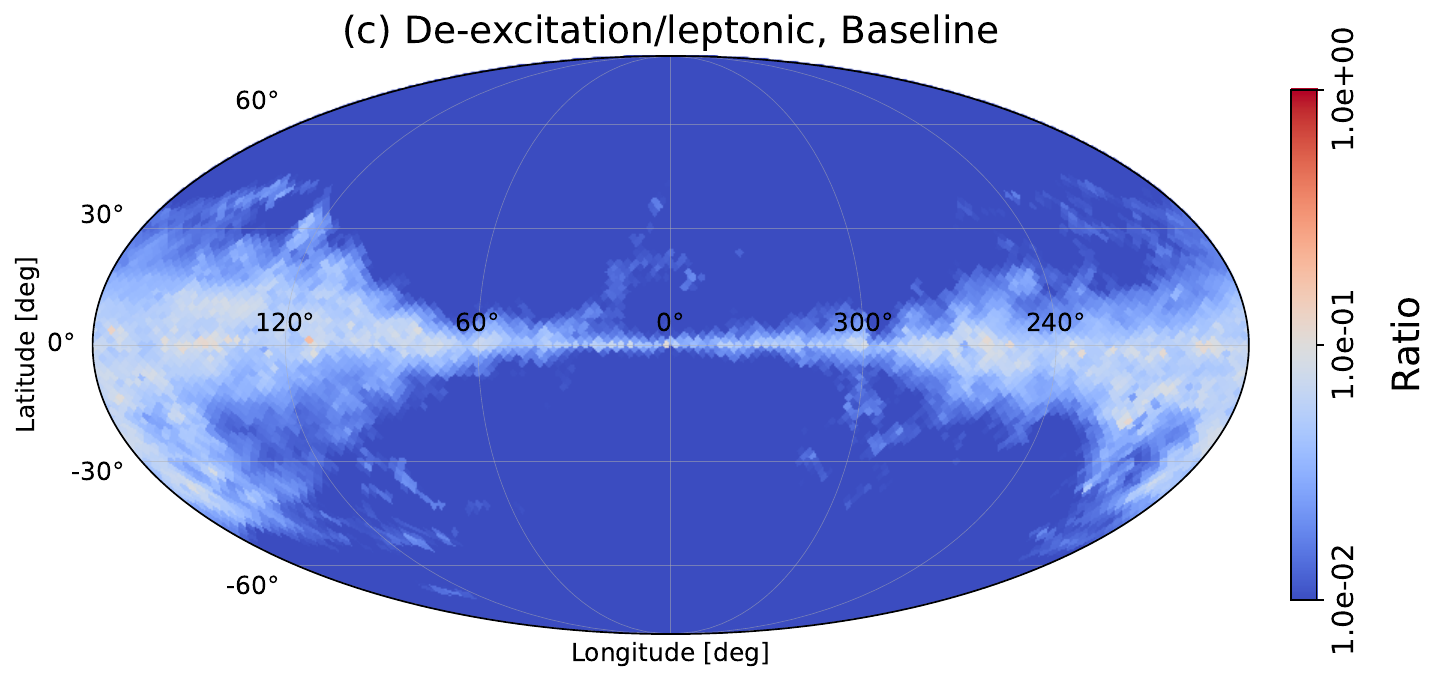}
\hfill
\includegraphics[width=0.48\linewidth,height=0.25\textheight,keepaspectratio]{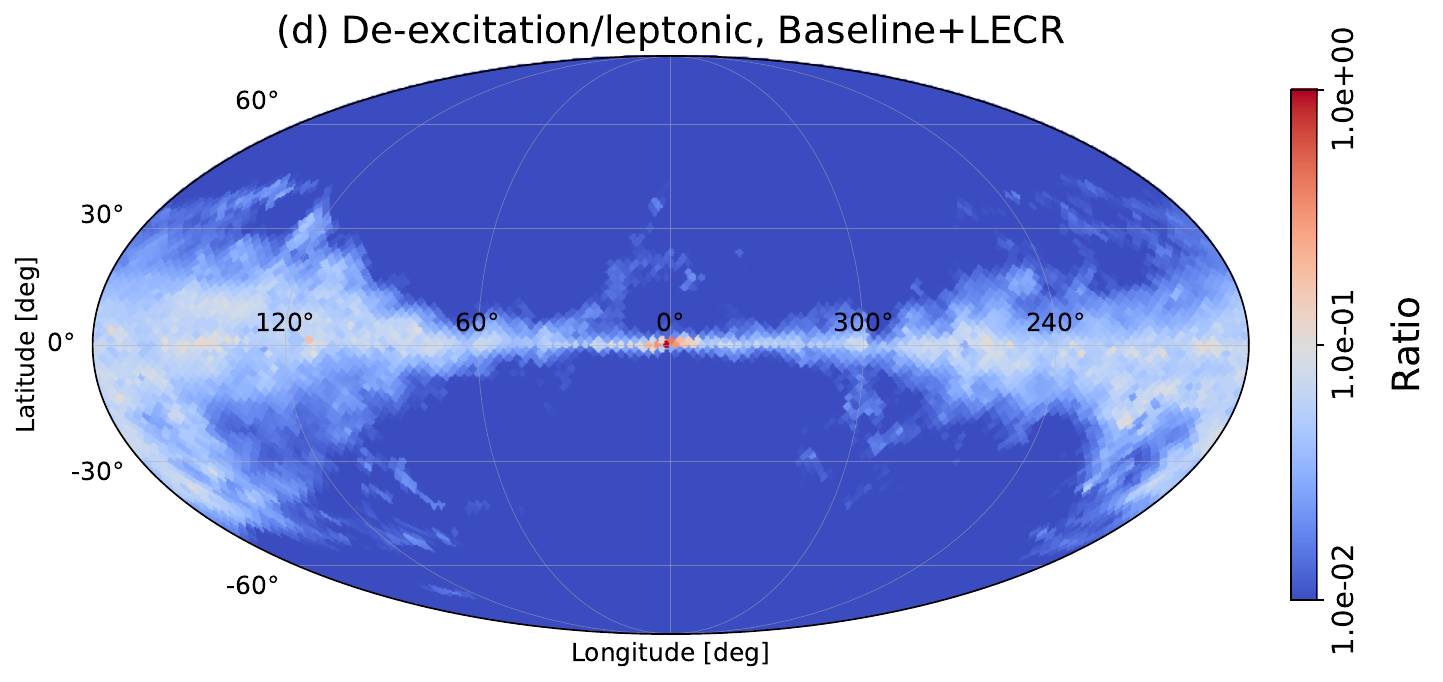}
\caption{
$\gamma$-ray flux maps at 4.4 MeV in Galactic coordinates.
Panels (a) and (b) show the de-excitation and leptonic-continuum emission for the baseline model, respectively, while panel (c) shows the ratio of de-excitation to leptonic-continuum flux for the baseline model.
Panel (d) shows the same as in panel (c), but for the combined baseline and fiducial LECR model.
}
\label{fig:skymap_comp}%
\end{figure*}

Fig.~\ref{fig:skymap_comp} compares the all-sky distributions of de-excitation and leptonic emission in the 4.4\,MeV band.
The de-excitation line emission traces the gas distribution (panel a), while the leptonic emission includes a substantial contribution from inverse Compton scattering in addition to electron bremsstrahlung, and thus depends not only on the gas density, but also on the spatial distribution of the interstellar radiation field (panel b).
Consequently, the ratio of de-excitation-line to leptonic-continuum emission is enhanced toward gas-rich regions.
However, for the baseline model, the de-excitation-line contribution remains modest, reaching at most ${\sim}20\%$ of the leptonic continuum.
Once the additional LECR component is included, the ratio of de-excitation-line to leptonic-continuum emission becomes significantly enhanced toward the GC, where it exceeds $100\%$ (panel d).

\begin{figure}
\centering
\includegraphics[width=\linewidth]{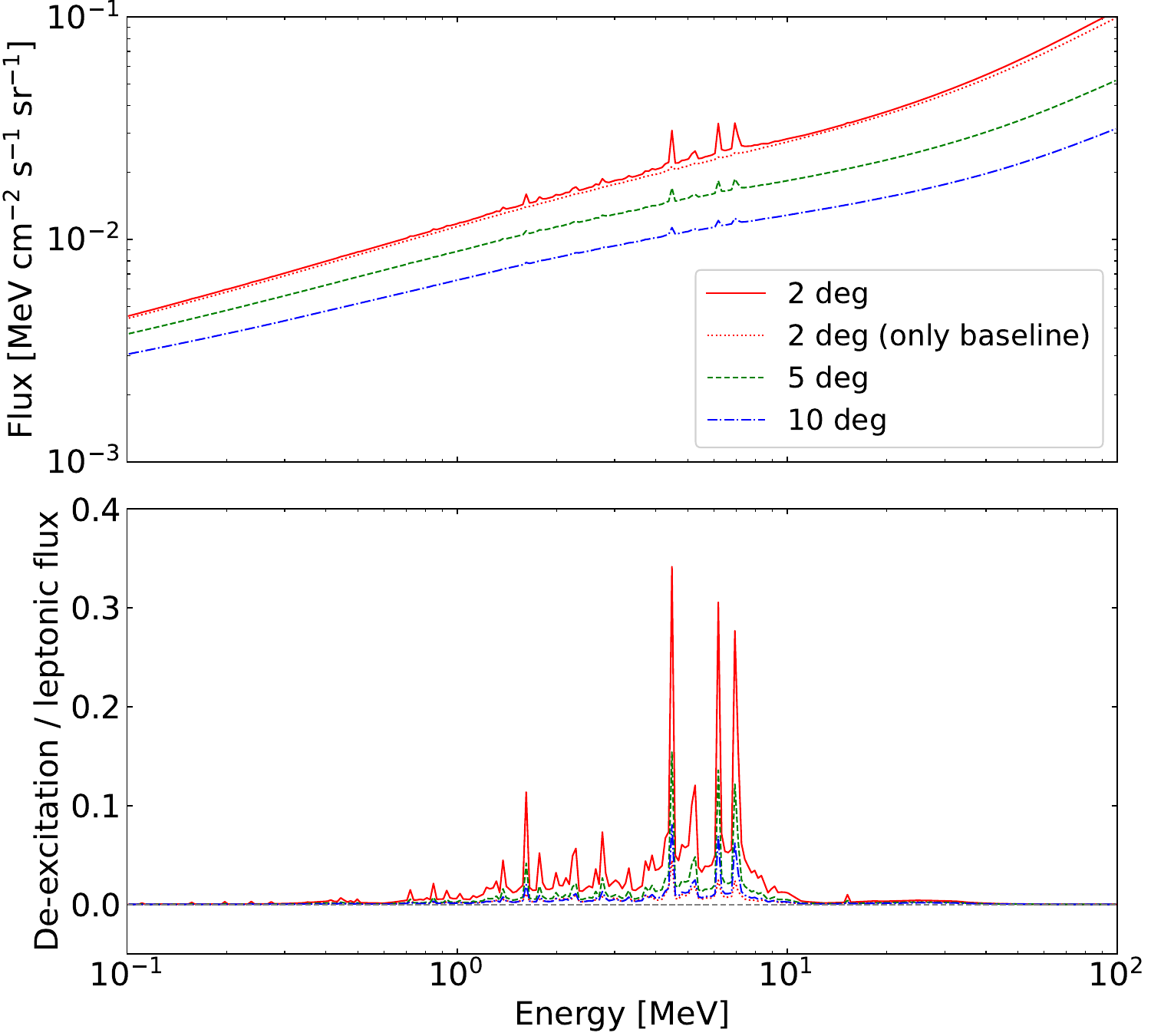}
\caption{$\gamma$-ray spectra integrated over circular regions of different radii centered on the GC.
The $\gamma$-ray spectra include contributions from de-excitation emission, inverse Compton scattering, electron bremsstrahlung, and neutral-pion decay.
The solid red, dashed green, and dot-dashed blue curves show the combined baseline and fiducial LECR model for integration radii of 2, 5, and 10$^{\circ}$, respectively, while the dotted red curve shows the baseline model for the 2$^{\circ}$ region.
The bottom panel shows the ratio of de-excitation to leptonic-continuum spectra.
}
\label{fig:spectrum_galcent}
\end{figure}

Fig.~\ref{fig:spectrum_galcent} shows the $\gamma$-ray spectra around the GC, integrated within circular regions of radius $2^{\circ}$, $5^{\circ}$, and $10^{\circ}$.
The $2^{\circ}$ angular scale is comparable to the source-localization accuracy of INTEGRAL/SPI \citep{Vedrenne2003A&A} and the angular resolution measure of the forthcoming MeV $\gamma$-ray observatory COSI \citep{Tomsick2024icrc}.
When the integration region is restricted to within $5^{\circ}$ of the GC, the ratio of de-excitation to leptonic emission exceeds $10\%$ at 4.4 and 6.1\,MeV.
For the central $2^{\circ}$ region, the 1.6\,MeV lines ($^{20}$Ne$^{*}$, $^{14}$N$^{*}$) and the 5.2--5.3\,MeV lines ($^{15}$N$^{*}$, $^{15}$O$^{*}$) also exceed $10\%$.
We note that observational results are also available for the $5^{\circ}$ integration region.
However, the statistical uncertainties remain too large to allow a detection of the de-excitation lines (see Appendix~\ref{sec:app:spectrum_vs_obs}).

\begin{figure}
\centering
\includegraphics[width=\linewidth]{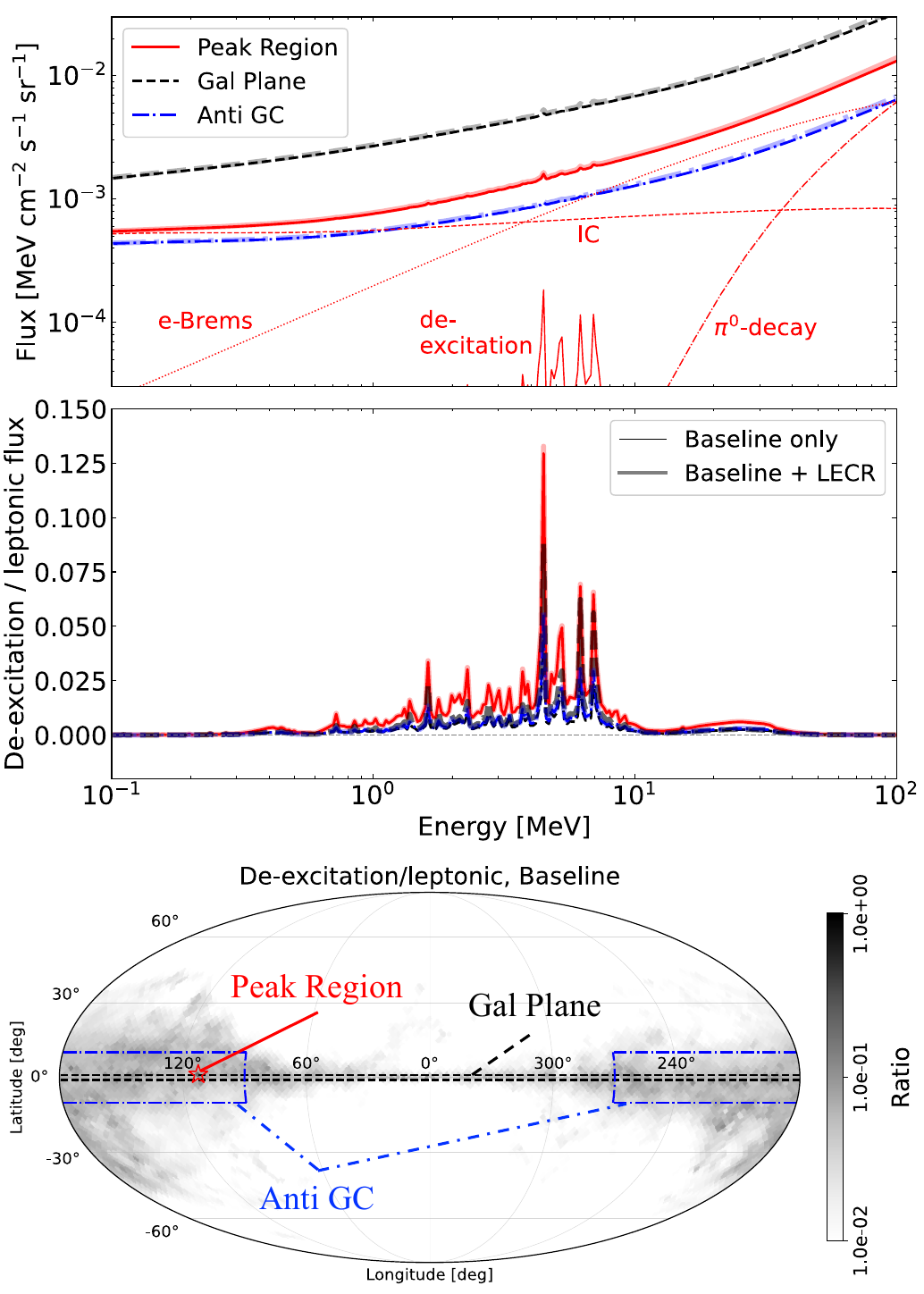}
\caption{
$\gamma$-ray spectra for the peak region (red solid), defined as a circular region within a radius of $1^{\circ}$ centered on the coordinates where the de-excitation-to-leptonic flux ratio is maximized in the baseline model, $(l,b)=(112.5^\circ, 1.2^\circ)$; the Galactic-plane region (black dashed; $|b|\le1^\circ$); and the anti-GC region (blue dot-dashed; $|l|\ge90^\circ$ and $|b|\le10^\circ$). 
The spectra include contributions from de-excitation emission, inverse Compton scattering, electron bremsstrahlung, and neutral-pion decay. 
For the peak region only, these individual emission components are also shown as thin red curves.
The thin and thick curves correspond to the baseline model and the combined baseline and fiducial LECR model, respectively. 
The middle panel shows the ratio of the de-excitation emission to the leptonic continuum.
The bottom panel illustrates the integration regions overlaid on the sky map shown in Fig.~\ref{fig:skymap_comp}(c).
}
\label{fig:spectrum_anti_galcent}
\end{figure}

We also search for alternative regions where de-excitation lines may be detectable. 
We extract $\gamma$-ray spectra from the following three regions: (i) a circular region of radius $1^{\circ}$ centered on the coordinates where the de-excitation-to-leptonic flux ratio is maximized in the baseline model, $(l,b)=(112.5^\circ, 1.2^\circ)$\footnote{Region (i) lies colse to a giant molecular cloud associated with the star-forming region NGC 7538 \citep{Ungerechts2000ApJ}, whose coordinates are $(l,b)=(111.5382^{\circ}, 0.8112^{\circ})$ \citep{Kharchenko2013A&A}.}; (ii) the Galactic-plane region ($|b|\le1^\circ$); and (iii) the anti-GC region ($|l|\ge90^\circ$ and $|b|\le10^\circ$). 
The resulting spectra are shown in Fig.~\ref{fig:spectrum_anti_galcent}. 
The ratio of the de-excitation emission to the leptonic continuum is highest in region (i), exceeding $10\%$ at the 4.4\,MeV line, while remaining below 10$\%$ in the other regions.
Including the additional LECR component does not significantly change these ratios, since its effect is largely confined to the vicinity of the GC.

\subsection{$\gamma$-ray line intensity ratio} \label{sec:results:line_ratio}

We examine how the LECR parameters affect the predicted de-excitation-line fluxes and their intensity ratios from different nuclei.
Although the absolute line fluxes depend on the LECR parameters, they are also affected by uncertainties in the gas density and elemental abundances.
As an alternative observable, we consider line-intensity ratios, which depend on the LECR spectrum because the production cross sections of individual $\gamma$-ray lines have different energy dependences (see Fig.~\ref{fig:csdata}).
Consequently, the relative contribution of the LECR component varies from line to line, making the intensity ratios sensitive probes of the LECR parameters.
In this work, we focus on the intensity ratio between the 6.1 and 4.4\,MeV de-excitation lines, which are expected to be the most readily detectable as shown in Fig.~\ref{fig:spectrum_galcent}.
The 4.4\,MeV line is primarily produced by $^{12}$C$^{*}$ via p+$^{12}$C and p+$^{16}$O reactions, whereas the 6.1\,MeV line originates from $^{16}$O$^{*}$ via p+$^{16}$O reactions.
The cross sections for p+$^{12}$C $\rightarrow$ $^{12}$C$^{*}$ and p+$^{16}$O $\rightarrow$ $^{16}$O$^{*}$ rise below 10\,MeV and peak at 10--20\,MeV as shown in Fig.~\ref{fig:csdata}, while the p+$^{16}$O $\rightarrow$ $^{12}$C$^{*}$ reaction peaks at higher energies around 25--30\,MeV.
The 4.4\,MeV line therefore receives a larger contribution from higher-energy CRs than the 6.1\,MeV line.
Consequently, the 6.1-to-4.4\,MeV intensity ratio increases as the LECR contribution becomes more significant.

\begin{figure*}
\includegraphics[width=\linewidth,keepaspectratio]{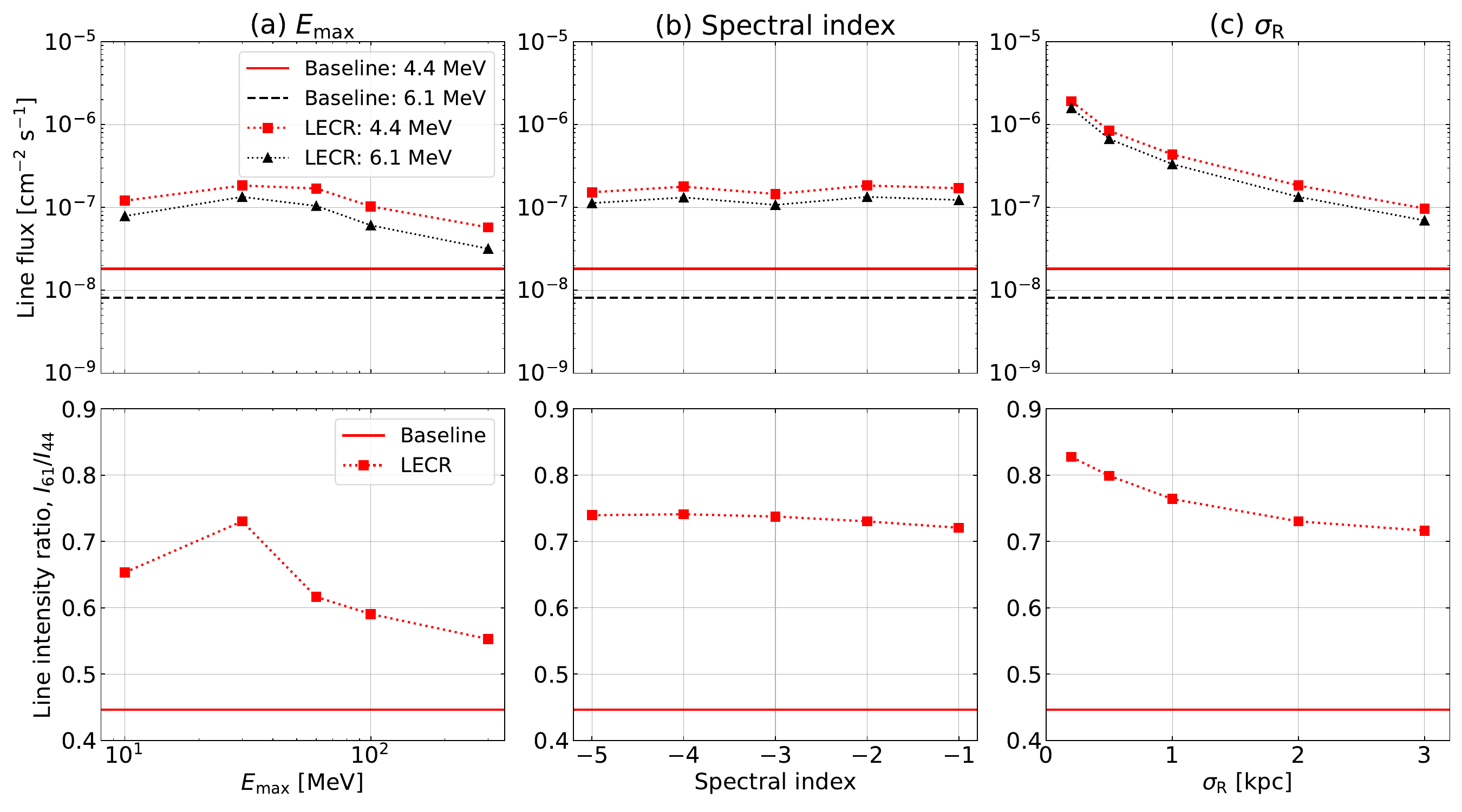}
\caption{Model dependence of the de-excitation-line fluxes at 4.4 and 6.1\,MeV (top panels) integrated within a radius of 2$^{\circ}$ from the GC and their ratios relative to the 4.4\,MeV line (bottom panels). 
Panels (a)--(c) show the dependence on $E_{\rm max}$, spectral index, and $\sigma_{\rm R}$, respectively.
In the top panels, the red squares and black triangles represent the 4.4 and 6.1\,MeV line fluxes for the additional LECR models, while the red solid and black dashed horizontal lines indicate the corresponding 4.4 and 6.1\,MeV intensities for the baseline model.
In the bottom panels, the square markers show the 6.1-to-4.4\,MeV line-intensity ratios, while the horizontal solid lines indicate the corresponding ratio for the baseline model.
}
\label{fig:line_intensity_ratio}
\end{figure*}

Fig.~\ref{fig:line_intensity_ratio} shows the de-excitation line fluxes at 4.4~MeV and 6.1\,MeV integrated within a radius of $2^{\circ}$ from the GC and their ratios relative to the 4.4 MeV line.
The line-flux ratio is 44.7$\%$ for the baseline model (44.7--45.7$\%$ across the SA0, SA50, and SA100 parameter sets), increasing to 73.0$\%$ in the fiducial LECR model.
The line-intensity ratio shows a relatively strong dependence on the assumed $E_\mathrm{max}$.
As $E_{\rm max}$ increases from 30 to 300\,MeV, the ratio decreases to 55.3$\%$, approaching the baseline-model value.
In contrast, the dependence on the spectral index is relatively weak: over the range $s=-1$ to $-5$, the line-intensity ratio varies by at most $3\%$.
This is likely because the effect of changing the spectral index on the LECR spectrum is limited to the energy range below the adopted maximum energy of $E_\mathrm{max}$=30\,MeV\,nuc$^{-1}$, leading to only a minor change in the de-excitation line intensity.

\section{Discussion} \label{sec:discussion}

\subsection{Model uncertainties} \label{sec:discussion:cross_section}

\begin{figure}
\centering
\includegraphics[width=\linewidth]{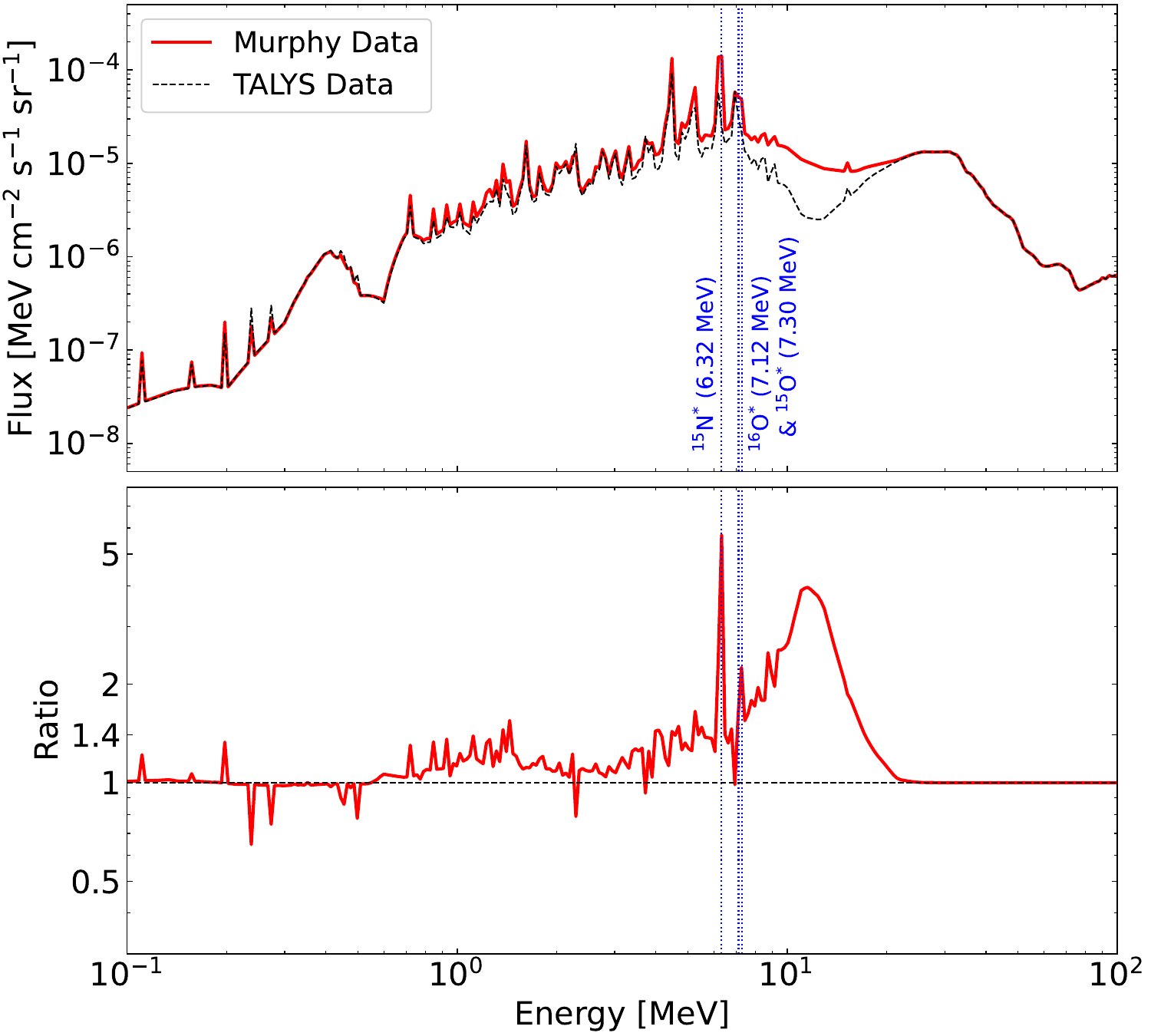}
\caption{
De-excitation $\gamma$-ray spectrum integrated over the GC region, $|l|\le30^{\circ}$ and $|b|\le15^{\circ}$, for the baseline model.
The red solid and black dashed curves show the spectra obtained with the cross-section data of \citet{Murphy2009ApJS} and TALYS, respectively.
The bottom panel shows the ratio of the spectrum calculated with the \citet{Murphy2009ApJS} cross sections to that calculated with TALYS.
}
\label{fig:spectrum_murphy}
\end{figure}

We discuss the uncertainties associated with the adopted nuclear reaction cross sections.
As described in Sect.~\ref{sec:method:cross_section}, all reaction cross sections adopted in this work were generated using TALYS.
Although the 4.4 and 6.1\,MeV transitions are generally reproduced well, comparisons with experimental data indicate discrepancies exceeding a factor of 2 above 100\,MeV.
To assess the impact of these uncertainties, we recalculate the $\gamma$-ray spectrum using the cross-section data provided by \citet{Murphy2009ApJS}, which are largely based on experimental measurements.
Fig.~\ref{fig:spectrum_murphy} compares the resulting spectrum with that obtained using TALYS.
Below ${\sim}5$\,MeV, the two calculations show good agreement within a factor of 1.5. 
In contrast, larger discrepancies are found around 6.3, 7.1, and 7.3\,MeV.
Due to the broad high-energy tails produced by inverse reactions, these differences propagate to higher energies and result in deviations of up to a factor of 4 around 10\,MeV.

\begin{figure}
\includegraphics[width=\linewidth,keepaspectratio]{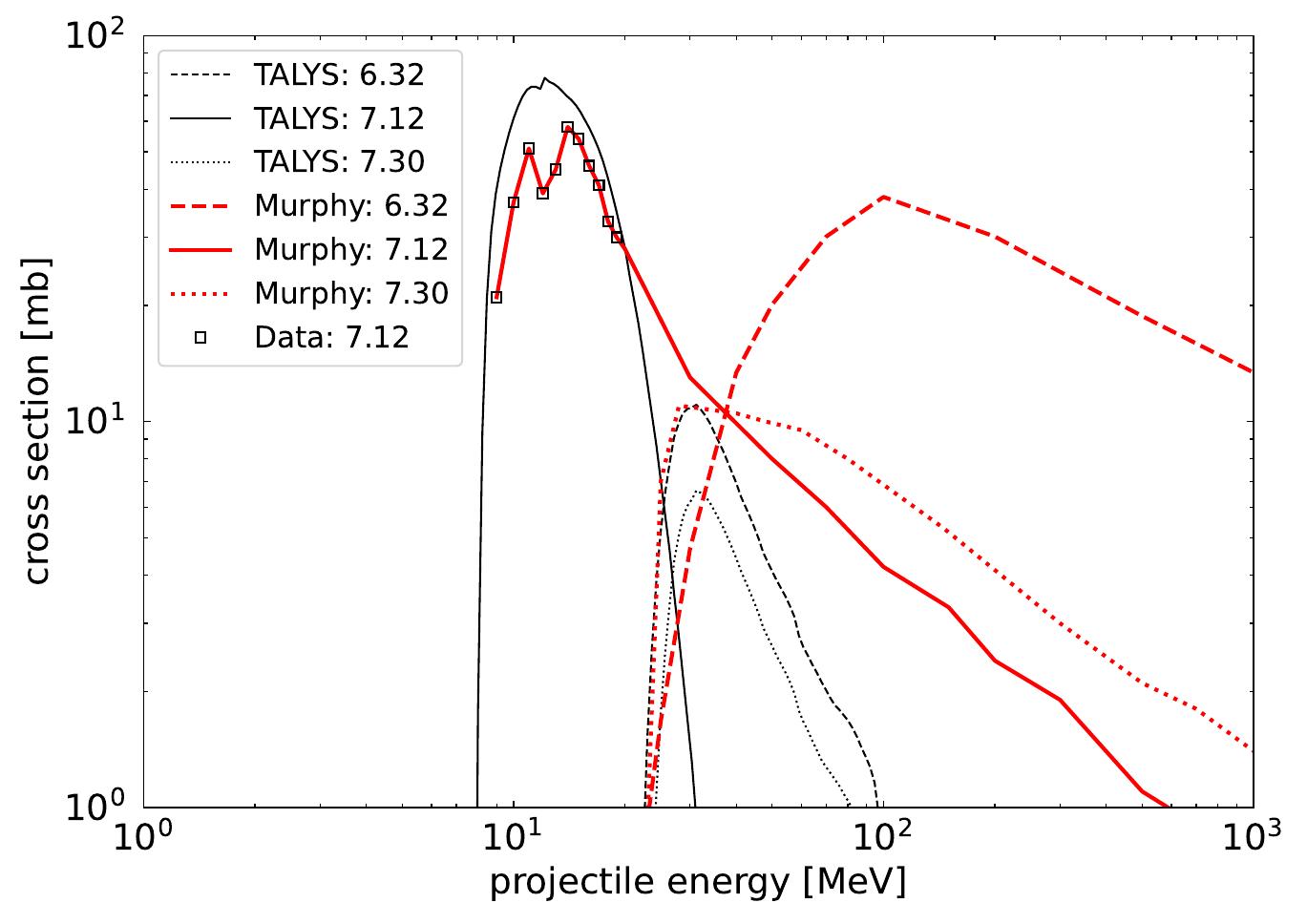}
\caption{
Cross sections of $\gamma$-ray line emission in p+$^{16}$O reactions. 
The red and black curves show the cross sections of \citet{Murphy2009ApJS} and TALYS, respectively.
The dashed, solid, and dotted curves correspond to the 6.32, 7.12, 7.30\,MeV $\gamma$-ray lines, respectively.
The open-square markers show experimental measurements from \citet{Kiener1998PhRvC}.
}
\label{fig:csdata_p16O_HE}
\end{figure}

The corresponding reaction cross sections are shown in Fig.~\ref{fig:csdata_p16O_HE}.
For the 7.1 MeV line, where experimental measurements are available, the TALYS and Murphy cross sections agree within a factor of 2 below 20\,MeV, while above 20\,MeV, the \citet{Murphy2009ApJS} data retains a substantial cross section whereas the TALYS calculation rapidly approaches zero.
Similar behaviour is found for the 6.3 and 7.3 MeV transitions, for which the \citet{Murphy2009ApJS} data expects significantly larger cross sections than TALYS.
Future laboratory measurements will therefore be important for reducing the cross-section uncertainties.
We note, however, that the impact of these uncertainties is limited to above 6.3 MeV and thus hardly affects the results and discussions regarding the 4.4 and 6.1\,MeV lines.

Note that the target-nuclei distribution, for which we adopt a simplified model as described in Sect.~\ref{sec:method:material}, may affect the absolute line fluxes; however, the 6.1-to-4.4\,MeV line-intensity ratio is primarily determined by the different energy dependences of the corresponding cross sections and therefore retains its sensitivity to the LECR parameters.


\subsection{Detectability with MeV $\gamma$-Ray Missions}

To date, observations in the MeV $\gamma$-ray band have been led primarily by CGRO/COMPTEL \citep{Schoenfelder1993ApJS} and INTEGRAL/SPI \citep{Vedrenne2003A&A}, neither of which has detected de-excitation line emission toward the GC \citep{Weinberger2021, Siegert2019}.
The upcoming mission, COSI \citep{Tomsick2024icrc}, employs a germanium-based Compton camera and is expected to achieve excellent energy resolution and line sensitivity ($3\times10^{-6}~\mathrm{cm^{-2}s^{-1}}$ for a 2-yr observation) in the MeV band.
The de-excitation line fluxes at 1.6, 4.4, and 6.1\,MeV, integrated within a radius of $2^{\circ}$ around the GC, can reach ${\gtrsim} 10^{-6}~\mathrm{cm^{-2}s^{-1}}$ for models with $\sigma_R \lesssim 0.2$\,kpc, which is comparable to the line sensitivity of COSI.
A more realistic assessment of the detectability will require simulations incorporating the instrument response and expected detector backgrounds, while recent advances in analysis techniques may further improve the sensitivity to weak de-excitation-line signals \citep[e.g.,][]{Yoneda2025A&A}.
In addition, as summarized by \citet{Aramaki2022arXiv}, many satellite missions have been proposed, including SMILE \citep{Tanimori2015ApJ}, e-ASTROGAM \citep{Tatischeff2016SPIE}, AMEGO-X \citep{Caputo2022JATIS}, GRAMS \citep{Aramaki2020APh}, GECCO \citep{Orlando2022JCAP}, and GammaTPC \citep{Shutt2025arXiv}.
These missions could provide improved line sensitivity below $10^{-6}~\mathrm{cm^{-2}s^{-1}}$, further enhancing the prospects for detecting de-excitation $\gamma$-ray emission and constraining the properties of the LECR population.

While we have focused on nuclear de-excitation emission as a component of the Galactic diffuse emission, previous studies have also investigated the prospects for detecting such emission from bright individual sources and regions, including the SNR Cas A \citep{Summa2011A&A, Liu2023MNRAS} and the GC \citep{Dogiel2009A&A, Liu2024PRD}.
The 4.4\,MeV line fluxes from both objects have been predicted to be at the $10^{-6}\,\mathrm{ph\,cm^{-2}\,s^{-1}}$ level.
Since Cas A is located away from the GC, where the Galactic diffuse de-excitation emission is relatively weak, its predicted flux is sufficiently high to be distinguishable from the diffuse component.
For the GC, our LECR model with $\sigma_\mathrm{R} = 0.2$\,kpc, which assumes an LECR distribution concentrated toward the GC and is thus conceptually similar to the scenarios considered in previous studies, yields a flux broadly comparable to previous estimates even after imposing the constraint from CR observations.

\subsection{Remarks on the additional LECR model} \label{sec:discussion:remark}

The additional LECR model capable of accounting for the observed ionization rates implies substantially higher CR power requirements.
The total power for injected CRs is estimated to be $1\times 10^{43}\,\mathrm{erg\,s^{-1}}$ for the fiducial case and $ 7\times 10^{42}\,\mathrm{erg\,s^{-1}}$ for the GC-concentrated case, with models having steeper spectral indices requiring even higher powers.
These values are about two orders of magnitude larger than the typical CR injection power of ${\sim}10^{41}\,\mathrm{erg\,s^{-1}}$ estimated for a standard Galactic CR model \citep{Strong2010ApJ}, corresponding to the baseline model considered in this work. 
They also exceed the CR power available from Galactic supernovae, $L_\mathrm{SNR} \sim 10^{41}\,\mathrm{erg\,s^{-1}}$, estimated assuming an explosion energy of $10^{51}$\,erg per supernova, a Galactic supernova rate of $0.03\,\mathrm{yr^{-1}}$, and a CR acceleration efficiency of $10\%$.
This fact again highlights the challenge of reconciling the ionization-rate measurements with the conventional SNR paradigm for Galactic CRs.

It should be noted that this study adopts a data-driven and spatially averaged treatment of CR injection and propagation to model $\gamma$-ray emission over the large regions relevant to Galactic diffuse emission.
In recent studies \citep[e.g.,][]{Phan2023PhRvD, Ravikularaman2025A&A}, more theoretically motivated models have been developed by adopting specific source geometries and solving CR transport, thereby providing a detailed treatment of the spatial variation of LECRs and the resulting ionization rates.
While a detailed investigation of such localized effects is beyond the scope of the present study, incorporating these spatial variations would refine predictions of $\gamma$-ray emission.

\section{Summary} \label{sec:summary}

Nuclear de-excitation $\gamma$-ray emission is a robust probe of the LECR flux, whose precise measurements would provide the missing link between astrophysics (CR acceleration and propagation) and astrochemistry (CR ionization).
In this work, we have established a 3D computational framework for predicting de-excitation $\gamma$-ray emission in the inner Galaxy by combining the CR propagation code GALPROP with the nuclear reaction code TALYS, enabling the construction of sky maps that can be directly compared with the Galactic diffuse continuum.

For standard CR propagation models, the predicted de-excitation emission is relatively weak, with the ratio of the de-excitation to leptonic continuum flux reaching at most $20\%$ over the entire sky at the 4.4\,MeV ($^{12}$C$^{*}$) band.
We further constructed models including an additional LECR component motivated by CR ionization rate measurements while remaining consistent with direct CR observations near Earth.
In these models, the de-excitation emission can exceed the leptonic continuum toward the GC.
In particular, for models with localized LECR source distributions ($\sigma_\mathrm{R} \lesssim 0.2$\,kpc), the predicted line fluxes integrated within $2^{\circ}$ of the GC can reach $\gtrsim 10^{-6}\,\mathrm{cm^{-2}s^{-1}}$ at 1.6, 4.4, and 6.1\,MeV, comparable to the expected sensitivity of the forthcoming MeV $\gamma$-ray observatory COSI.
We also investigated the diagnostic potential of de-excitation line-intensity ratios.
The 6.1-to-4.4\,MeV line ratio is particularly sensitive to the maximum energy of the injected LECR spectrum ($E_\mathrm{max}$), increasing by up to ${\sim}30\%$ relative to that ($45\%$) of the baseline model.

Our results demonstrate that future MeV $\gamma$-ray observations will be capable of testing the existence of an additional Galactic LECR component and placing robust constraints on its population and properties.

\begin{acknowledgements}
Tomohiko Oka acknowledges support by JSPS Overseas Research Fellowships.
Saurabh Mittal acknowledges support by the {Bundesministerium für Wirtschaft und Energie via the Deutsches Zentrum für Luft-und Raumfahrt (DLR)} under Contract No. {50\,OO\,2219}.
Laura Eisenberger acknowledges support by the Bundesministerium für Wirtschaft und Energie via the Deutsches Zentrum für Luft-und Raumfahrt (DLR) under Contract No. 50OR2413 and is grateful for the support of the Studienstiftung des Deutschen Volkes.
Dimitris Tsatsis and Tristan Bouchet acknowledge support from the DFG/LIS project SI2502/6-1, project number 551127478.
Rudi Reinhardt acknowledges support by the Bundesministerium für Forschung, Technologie und Raumfahrt as part of the FORKA (Forschung für den Rückbau kerntechnischer Anlagen) funding program under the reference number 15S9455A-E.
Hiroki Yoneda is supported by JSPS KAKENHI Grant Number 23K13136 and 26H00833.
\end{acknowledgements}

\bibliographystyle{aa}
\bibliography{main}

\begin{appendix}

\section{Additional comparisons of predicted $\gamma$-ray spectra with observations} 
\label{sec:app:spectrum_vs_obs}

\begin{figure*}
\includegraphics[width=0.48\linewidth,height=0.25\textheight,keepaspectratio]{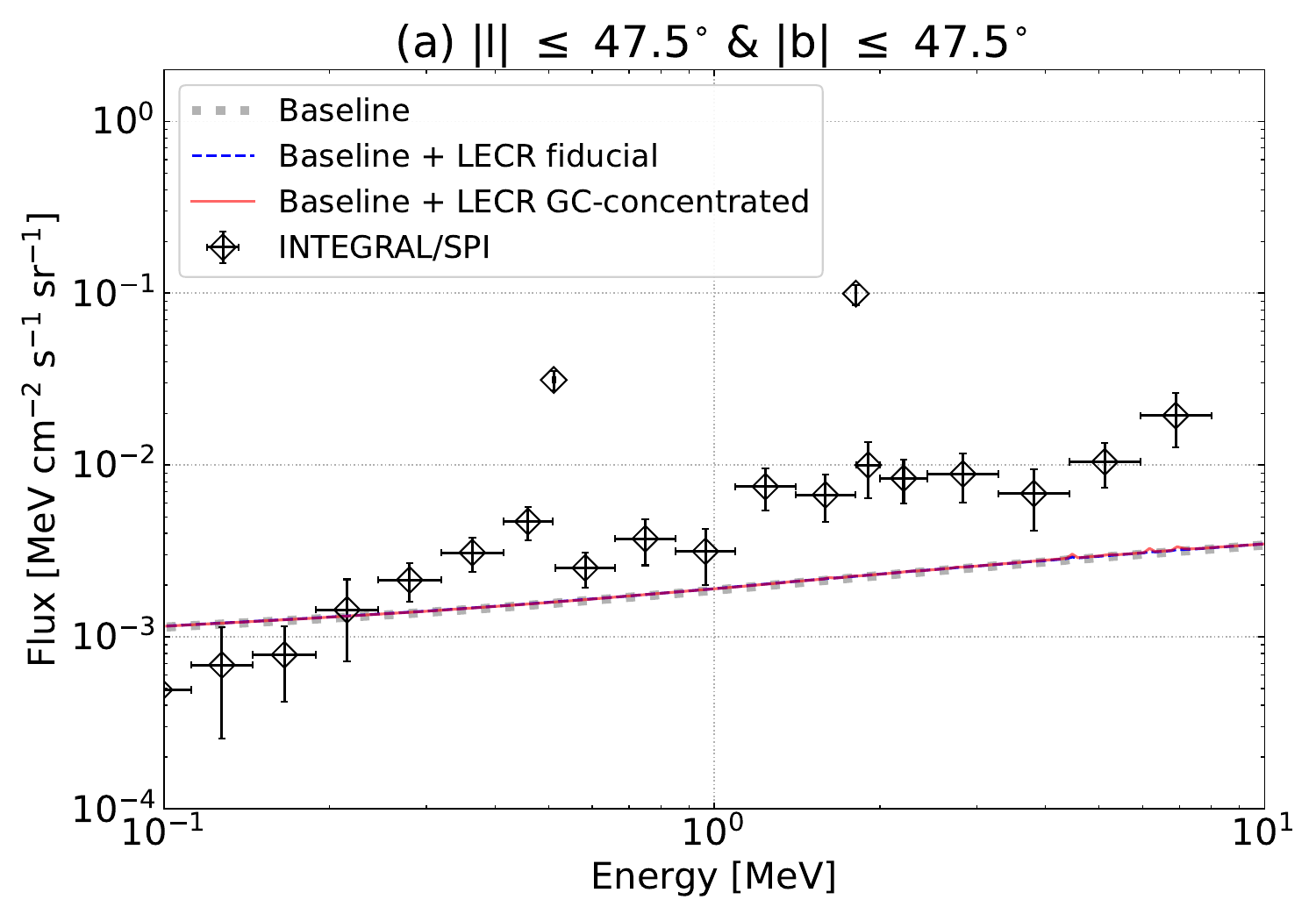}
\hfill
\includegraphics[width=0.48\linewidth,height=0.25\textheight,keepaspectratio]{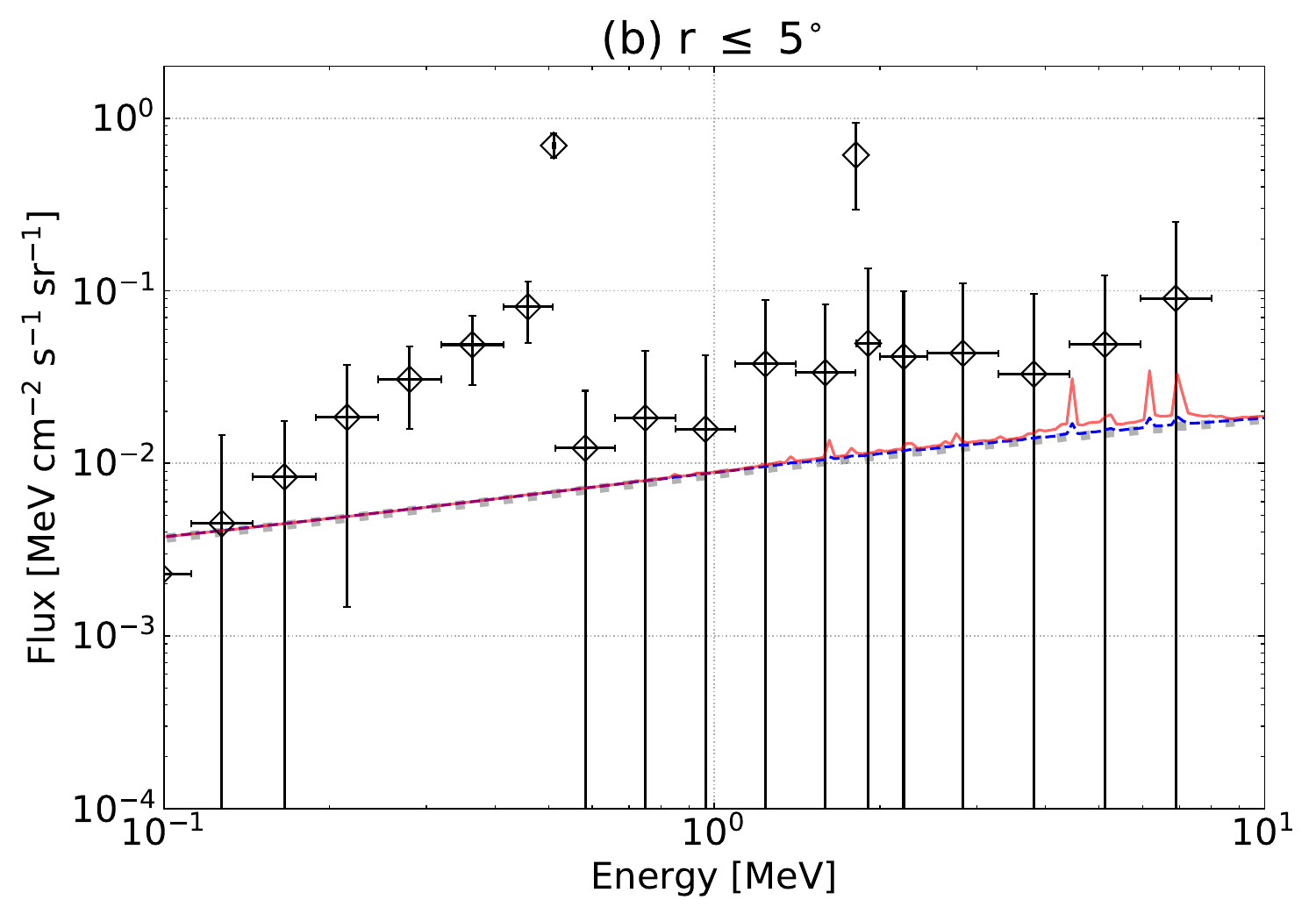}
\includegraphics[width=0.48\linewidth,height=0.25\textheight,keepaspectratio]{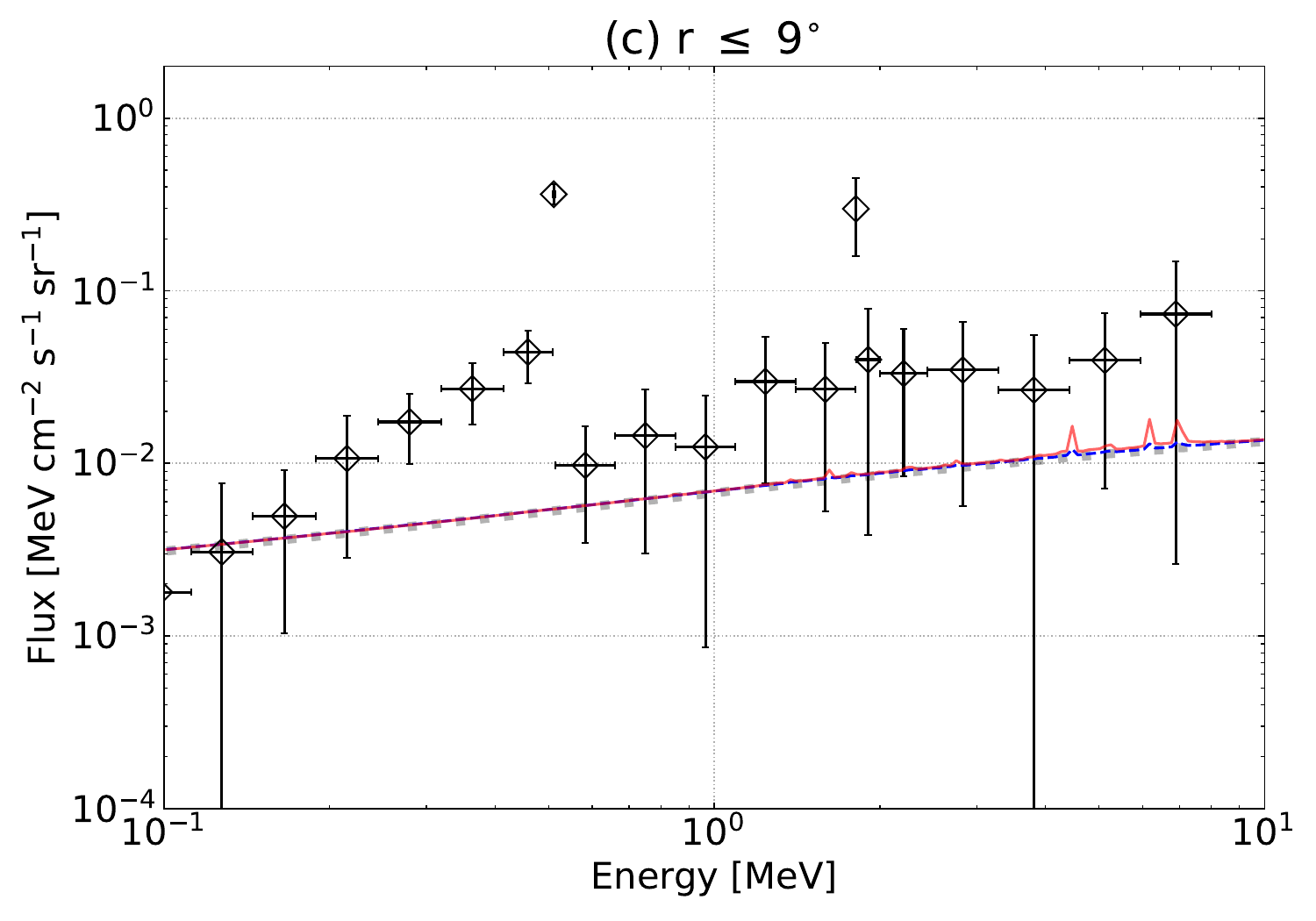}
\hfill
\includegraphics[width=0.48\linewidth,height=0.25\textheight,keepaspectratio]{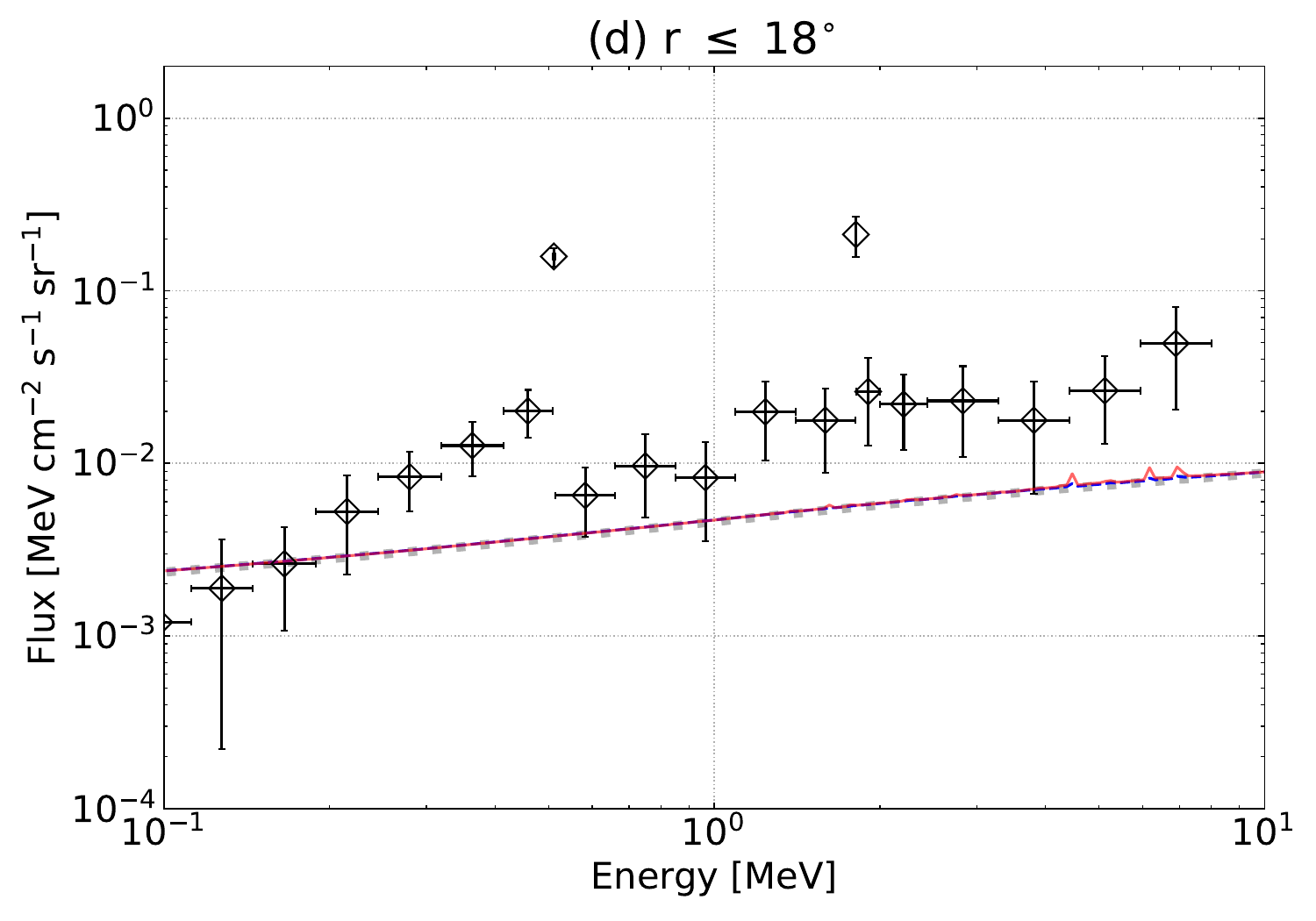}
\caption{
$\gamma$-ray spectra integrated over (a) $|l| \leq 47.5^{\circ}$ and $|b| \leq 47.5^{\circ}$, and circular regions centered on the GC with radii of (b) 5$^{\circ}$, (c) 9$^{\circ}$, and  (d) 18$^{\circ}$.
The black dotted curves show the baseline model, while blue dashed and red solid curves show the combined spectra of the baseline model with the fiducial and GC-concentrated LECR model, respectively.
In the latter, $\sigma_\mathrm{R}=0.2$\,kpc is adopted instead of the fiducial value of 2\,kpc.
The data points show the INTEGRAL/SPI measurements from \citet{Berteaud2022PhRvD} and \citet{Siegert2022A&A} for panel (a), and \citet{Das2025arXiv} for panels (b--d).}
\label{fig:spectrum_vs_obs}%
\end{figure*}

Fig.~\ref{fig:spectrum_vs_obs} shows the $\gamma$-ray spectra integrated over different regions for which observational results are available \citep{Berteaud2022PhRvD, Siegert2022A&A, Das2025arXiv}.
As in Fig.~\ref{fig:spectrum_galcent}, the predicted de-excitation lines become more prominent for smaller integration regions.
However, the statistical uncertainties of the observed spectra also increase.
Consequently, all models remain consistent with the observed spectra and the non-detection of de-excitation lines.

\end{appendix}

\end{document}